\documentclass[aps,
prd,
english,
nofootinbib,
longbibliography,
showkeys,
reprint
]{revtex4-2}

\pdfoutput=1
\usepackage{amsfonts,amsmath,amssymb}
\usepackage{graphicx}
\usepackage[utf8]{inputenc}
\usepackage{hyperref}
\usepackage{babel}
\usepackage{listings}
\usepackage{matlab-prettifier}
\usepackage{enumitem}
\usepackage{lipsum}

\newcommand\e{{\rm e}}

\newcommand\be{\begin{equation}}
\newcommand\ee{\end{equation}}
\newcommand\bea{\begin{eqnarray}}
\newcommand\eea{\end{eqnarray}}

\begin{document}

\def\rhoo{\rho_{_0}\!} 
\def\rhooo{\rho_{_{0,0}}\!} 

\begin{flushright}
\phantom{
{\tt arXiv:2006.$\_\_\_\_$}
}
\end{flushright}

{\flushleft\vskip-1.4cm\vbox{\includegraphics[width=1.15in]{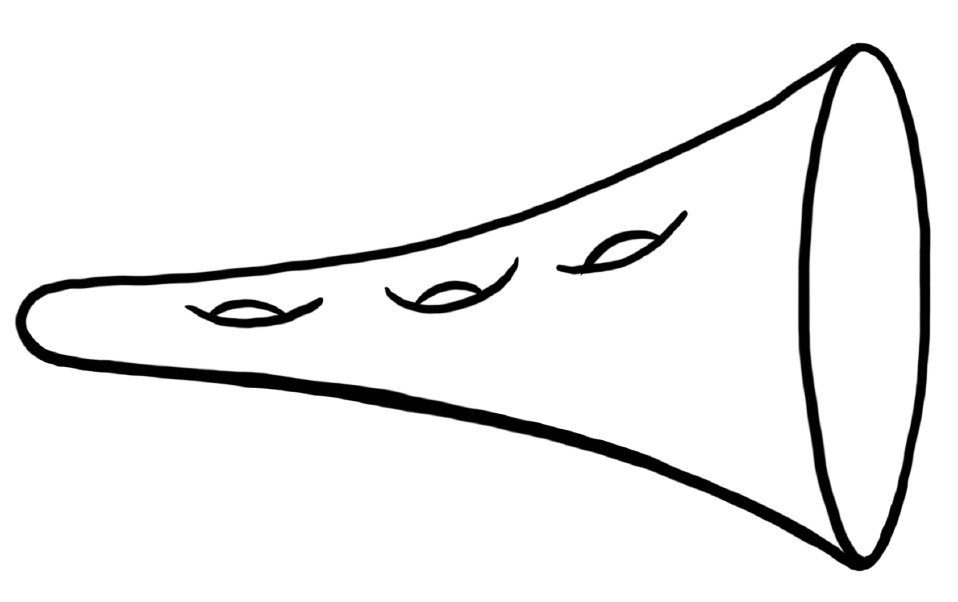}}}

\title{
  Universal formulae for correlators of a broad class of models}
\author{Clifford V. Johnson}
\email{cliffordjohnson@ucsb.edu}

\affiliation{Department of Physics, Broida Hall,   University of California, 
Santa Barbara, CA 93106, U.S.A.}


\begin{abstract}

A simple method is presented for 
deriving universal formulae for the correlators, frequently denoted $W_{g,n}(\{z_i\}),  i=1,..n$, of a wide range of models of physical and mathematical interest. 
While many alternative methods exist for constructing such correlators, these formulae can be simply written in terms  of one defining function and its derivatives. The method has been  applied to the Airy and Bessel models, various minimal string and superstring theories, and their associated intersection theory settings, ordinary and various kinds of supersymmetric Weil-Petersson volumes,  and more besides.  For all these cases, their $W_{g,n}(\{z_i\})$ are just all specializations of the {\it same} universal formulae. 
A special variant of the method useful for ${\cal N}{=}1$ supersymmetric cases is also presented. It allows for  swift derivations of   Norbury's three closed-form formulae for the volumes $V_{g,n}$ ($g{=}1,2,3$) of ${\cal N}{=}1$ supersymmetric Weil-Petersson volumes, and generalizations of them to a wider set of models. Moreover a new  closed-form formula for the genus 4 case $V_{4,n}$ is derived. The straightforward method for how to derive such  formulae  for $g{>}4$ cases is described. Throughout,  crucial roles are played by the underlying integrable  KdV flows, as well as the Gel'fand-Dikii 
equation.

\end{abstract}


\maketitle


\section{Introduction}
\label{sec:introduction}

\noindent This paper will be concerned with a large class of theories  that have an energy  spectrum  characterized by the continuous   density of states $\rho(E)$ (here $E$ is energy) (possibly with a $\delta$-function discrete component as well).  It has  an expansion in a small parameter  $\hbar$, and the leading part is  $\rho_0(E)$. The continuous part begins above some threshold energy~$E_0$. While various random matrix models serve as  prototypes, in mind is a broader list of systems from quantum gravity and black hole physics, string theory, quantum chaos, and statistical physics, as well as various related mathematical problems in  areas such as random geometry,  enumerative geometry, and integrable systems.

Often, the models emerge as a limit of a system with~$N$  energy levels, and ``leading'' means large~$N$, with the  expansion parameter being $\hbar{\sim}1/N$. The  generic energy levels become  dense on an interval of fixed size, and hence a continuous function  can be used to describe them well.  The spectral density~$\rho(E)$ arises in  certain scaling limits that extract key universal physics.
Importantly, an underlying integrable (Korteweg-de Vries (KdV)) system plays a role in  characterizing the physics, although the physical dynamics itself is typically chaotic. 

There is an integral representation of the leading density in terms of another function:
\begin{equation}
    \label{eq:integral-representation}
    \rho_0(E)=\frac{1}{2\pi\hbar}\int_{-\infty}^\mu\frac{\Theta(E-u_0(x)) dx}{\sqrt{E-u_0(x)}}\ ,
\end{equation}
taking as input the function $u_0(x)$ on $x$, the real line. Also,   $u_0(\mu){=}E_0$, which can be taken as a definition of the parameter $\mu$. Different  models, and hence their distinct leading spectral densities,  correspond to different $u_0(x)$. 

An example  physical setting  for all of this, as already mentioned, is the behaviour of certain large~$N$ random matrix models in a ``double-scaling limit'' (DSL)~\cite{Brezin:1990rb,Douglas:1990ve,Gross:1990vs}, originally motivated by formulating  sums over two dimensional Euclidean geometries in order to get to grips with worldsheet formulations of string theory, and $D{=}2$ quantum gravity more generally.  The large $N$ expansion in this case can also be interpreted~\cite{'tHooft:1973jz,Brezin:1978sv,Bessis:1980ss} as an expansion in the topology of the  surfaces. Surfaces with $g$ handles come with a factor $N^{2-2g}$. In the DSL, $1/N$ gets renormalized into the  parameter~$\hbar$. So perturbation theory is in powers of $\hbar$, with contributions organized by genus.\footnote{Random matrix models in the DSL are of wider physics interest than just 2D gravity and string theory. There is a wealth of interesting statistical physics phenomena to be found in this same  regime, as well as connections to other fields such as quantum chaos. See {\it e.g.}  ref.~\cite{forrester2010log} for more.}

It is natural to want to compute $n$-point correlations $\langle\rho_0(E_1)\cdots\rho_0(E_n)\rangle_0$, as well as corrections to $\rho_0(E)$ and all such correlations as an (asymptotic) expansion in $\hbar$. The central result of this paper is a method for deriving a {\it universal  formula} for these corrections (denoted~${\widetilde W}_{g,n}$), very simply, at any $g$ and any~$n$, written purely in terms of the function $u_0(x)$ and its derivatives, evaluated at $x{=}\mu$. Moreover, for a class of  ${\cal N}{=}1$ supersymmetric cases, the method readily yields, for a given $g$, an even simpler {\it closed form} all-$n$ formula!

Note that the approach of this paper is {\it not} equivalent to certain other  methods for obtaining these correlators, such as the Chekhov-Eynard-Orantin ``topological recursion'' (TR)~\cite{Chekhov:2006vd,Eynard:2007kz}, or, in the case when the $W_{g,n}$ are Laplace transforms of the volumes~$V_{g,n}$ of  the compactified moduli space of Riemann surfaces  with $n$ geodesic boundaries and genus~$g$, the  recursive approach of Mirzakhani\cite{Mirzakhani:2006fta}\footnote{For this geometrical problem, Eynard and Orantin\cite{Eynard:2007fi} showed the equivalence between topological recursion and Mirzakhani's relation (after Laplace transform).} with supersymmetric generalizations by Stanford and Witten (${\cal N}{=}1$)\cite{Stanford:2019vob} and Turiaci and Witten (${\cal N}{=}2$) \cite{Turiaci:2023jfa}. 
A key reason is that those recursion relations are structured according to the different ways of embedding a distinguished ``pair-of-pants''  into  a surface $\Sigma_{g,n}$ such that it decomposes into a number of smaller surfaces $\Sigma_{g',n'}$, (with $2g'{-}2{+}n'{<}2g{-}2{+}n$). All of the amplitudes $W_{g',n'}$ must be computed in order to determine~$W_{g,n}$. 
In this paper, however, constructing $W_{g,n}$ (and hence $V_{g,n}$ where appropriate) is  a more direct affair by comparison. The method uses the fact that there is an underlying organization by the integrable Korteweg-de Vries (KdV) flows, with initial conditions given by an ordinary differential  equation (ODE) called a ``string equation'', whose (genus) perturbative structure  has a quite general form across many models. The genus expansion, {\it e.g.,} for $W_{g,1}$, directly emerges from simple recursive expansion of the Gel'fand-Dikii ODE~\cite{Johnson:2024bue}. The key observation in this paper is that when applied judiciously, the operator for adding boundaries simplifies to an elementary operation that is easy to iterate, readily yielding quite general formulae for $W_{g,n}$ that apply across many models, bosonic and supersymmetric.\footnote{In principle, this paper could  have been written in 1991, since the basic ingredients were all in place by then! But of course  the many subsequent discoveries that motivate its writing now were not yet in place. Since then  there has appeared  a vast literature exploring the connections between integrable hierarchies, matrix models, and geometry, but the simple method outlined in this paper, and the  formulae for $W_{g,n}$ that result, do not seem to have appeared in print  (at least as searches have revealed  so far).  The same appears to be true for the  observed  universality of the formulae across bosonic and supersymmetric examples.}

So what form do the formulae take? A well known example is the simple result~\cite{Ambjorn:1990ji,Moore:1991ir,Ginsparg:1993is} from long ago for the genus zero $n$-point correlator 
$\langle{\widetilde W(\ell_1)}\cdots {\widetilde W(\ell_n)}\rangle$
(for $n{\geq}2$) of ${\widetilde W(\ell)}$, the Laplace transform of $\rho(E)$, {\it i.e.,} the loop of length $\ell$ \cite{Ginsparg:1993is,Gross:1990aw,Banks:1990df}:
\begin{eqnarray}
\label{eq:W0n}
\hskip-0.5cm 
    {\widetilde W}_{0,n}(\{(\ell_i)\})
   =\frac{\sqrt{\ell_1\cdots\ell_n}}{2\pi^\frac{n}{2}\ell_T}\left(\frac{\partial}{\partial x}\right)^{n-2}\!\!{\rm e}^{-\ell_T\,u_0(x)}\Biggr|_{x=\mu}\!
   \ ,
\end{eqnarray}
where $\ell_T{=}\sum_{i=1}^n\ell_i$. This paper will show how to swiftly construct analogues of this formula for other $g$ and $n$.

Such formulae are nice to have, of course, but there is additional value here because these same kinds of  correlators (in this general class) have other interpretations in problems to be found in other  areas of physics, as well as  beyond to various fields of Mathematics. The analogues of these correlators 
are sometimes far more difficult to compute in their native setting, or harder to write general formulae for. So the purpose of the paper is to present the method, show some examples, and also try to highlight some of the connections and uses of the framework.

The best  motivation is probably to just jump right in with  some results. Consider the formula:\cite{Johnson:2024bue}
\begin{eqnarray}
\label{eq:W11-E}
 &&\hskip-1.0cm{\widetilde W}_{1,1}(E) =    \frac{u_0^\prime(\mu)}{32[u_0(\mu)-E]^\frac{5}{2}}-\frac{u_0^{\prime\prime}(\mu)}{48 u_0^\prime(\mu)[u_0(\mu)-E]^{\frac32}}\ ,
\end{eqnarray}
or its Laplace transform (exchanging $E$ domain for~$\ell$ domain):
\begin{equation}
\label{eq:W11-ell}
\hskip-1.0cm{\widetilde W}_{1,1}(\ell)=\frac{1}{24}
\sqrt{\frac{\ell}{\pi}}\, {\mathrm e}^{-\ell {u_0}(\mu)} 
\left(
\ell u^\prime_0(\mu)-\frac{u_0^{\prime\prime}(\mu)}{u_0^\prime(\mu)} 
\right)\ .
\end{equation}
In fact (\ref{eq:W11-E}) encapsulates the first correction to the spectral density, and in the form~(\ref{eq:W11-ell}) it is the first in a family of genus 1 generalizations of~(\ref{eq:W0n}).\footnote{In the conventions of this paper, ${\widetilde W}_{1,1}(E)$ naturally comes at order  $\hbar^2$, and more generally ${\widetilde W}_{g,n}(E)$ will come with $\hbar^{2g}$. Converting to the $n$-point result in a $\rho(E)$ correlator will involve dividing by $\hbar^n$.}

Here are two illustrative examples of models we can employ through the course of  the paper to show the use of the formulae. First is the case $u_0(x){=}{-}2x$, corresponding to  the universal model living at the edge of Wigner's semi-circle, arising in the Gaussian unitary ensemble of random matrices. This edge limit is known   as the Airy model. It turns out that $\mu{=}0$ here, so $u_0(0){=}0$ and $u_0^\prime(0){=}{-}2$, and (after continuing $E{\to}\,{-}E$ and dividing by $\pi\hbar$) the resulting ${\widetilde W}_{1,1}(E){=}{-}1/(16 E^\frac52)$ is the first (well-known) correction  to the model's leading spectral density $\rho_0(E){=}\sqrt{E}/\hbar\pi$.

On the other hand, this is also a result in Kontsevich-Witten intersection theory and topological gravity~\cite{Witten:1989ig,Witten:1990hr,Kontsevich:1992ti}. (See ref.~\cite{Eynard:2016yaa} for an excellent  review of the subject in the context of the $W_{g,n}$ correlators.) To appreciate this and some results to come, change variables from $E$ to $z$ using  $z^2{=}u_0(\mu){-}E$ and convert from  the one-form ${\widetilde W_{1,1}(E)dE}$,   to $W_{1,1}(z)dz$, to get (remember to multiply equation~(\ref{eq:W11-E}) by Jacobian $-2z$): 
\begin{eqnarray}
\label{eq:W11-z}
    \hskip -0.3cm W_{1,1}(z)
    =\frac{A_{11}}{z^2}+\frac{B_{11}}{z^4}
    =\frac{1}{24}\frac{u_0^{\prime\prime}(\mu)}{u_0^\prime(\mu)}\frac{1}{z^2}-\frac{u_0^\prime(\mu)}{16}  \frac{1}{z^4}\ ,
\end{eqnarray}
(where $A_{11}$ and $B_{11}$ are defined for later reference) 
and the result for Airy is $W_{1,1}^{\rm Airy}(z){=}\frac{1}{8z^4}$. (Note here, and henceforth, that when written in terms the~$z$ variable, with Jacobian factor absorbed, we do not have a tilde on the $W$.)
This object captures a topological invariant, and is written $W_{1,1}(z){=}\langle \tau_1\rangle_1\frac{3!!}{z^4}$. After stripping off the combinatorial  factor of\footnote{\label{fn:combinatorics-note}The combinatorial factor is $(2d+1)!!$ where $d$ is the power of $z$ in the denominator, and $d=3g-3+1$.}~$3!!{=}3$, the topological quantity is $\langle \tau_1\rangle_1{=}1/24$, the intersection number obtained by integrating a single power (in this case) of the $\psi$-class on $\overline{{\cal M}}_{1,1}$, (a 
compactification of) the moduli space of Riemann surfaces of genus $1$ with one puncture. ($\tau_m$ means integrate the $m$th power of the $\psi$-class inserted at a point.)
The full family of  objects $W_{g,n}(\{z_i\})$  pertains to the  cases of multiple ($n$) punctures, on genus $g$. The numbers encode intersection numbers of powers of $\psi$--classes again.\footnote{We will not review that here, but instead point the reader to the ample introductory literature. See for example refs.~\cite{Wolpert2010WeilPetersson,Eynard:2016yaa,Bouchard:2024fih}.}

The second example has $u_0(x)$ defined by~\cite{Saad:2019lba,Okuyama:2019xbv}: \begin{equation}
\label{eq:JT-WP-u0-equation}
\sqrt{u_0(x)}I_0(2\pi\sqrt{u_0(x)}){+}x{=}0\ ,
\end{equation}
from which can be readily extracted the following behaviour   at $x{=}\mu{=}0$: \begin{eqnarray}
\label{eq:JT-Wp-u0-derivatives}
 &&u_0(0)=0\ ,\quad u_0^\prime(0)={-}2\ ,\quad u_0^{\prime\prime}(0)={-}4\pi^2\ ,\quad  \\ \nonumber
 &&u_0^{(3)}(0)=-20\pi^4\ ,\quad u_0^{(4)}(0)=-\frac{488\pi^6}{3}\ ,\\ \nonumber
 &&u_0^{(5)}(0)=-\frac{5516\pi^8}{3}\ ,\quad u_0^{(6)}(0)=-\frac{399568\pi^{10}}{15}\ ,
\end{eqnarray}
collected here 
since we will need them later. (Here $u_0^{(p)}$ means the $p$th $x$-derivative of $u_0(x)$.) Putting this all into our equation~(\ref{eq:W11-E}), the resulting $\widetilde{W}_{1,1}(E)$ is again a correction to a matrix model, this time it captures~\cite{Saad:2019lba}  Jackiw-Teitelboim (JT) gravity~\cite{Jackiw:1984je,Teitelboim:1983ux}, and this is the first correction to the spectral density $\rho_0(E){=}\sinh(2\pi\sqrt{E})/4\pi^2\hbar$. On the other hand using the form  given in equation~(\ref{eq:W11-z}):
\begin{equation}
\label{eq:W11-V11-JT}
    W_{1,1}(z)=\frac{\pi^2}{12z^2}+\frac{1}{8z^4}
    =    \int_0^\infty \! bdb\,\e^{-b z}\,  V_{1,1}(b)\ ,
\end{equation}
where the nice object coming from the Laplace transform:
\begin{equation}
\label{eq:V11-JT}
    V_{1,1}(b)=\frac{(4\pi^2+b^2)}{48}\ ,\quad\text{(JT/WP)}
\quad
\end{equation}is  the Weil-Petersson volume of the moduli space  of bordered Riemann surfaces of  genus 1 with a single geodesic boundary of length $b$. This hence connects to  problems of keen interest in Mathematics~\cite{penner1992weil,Zograf1993WeilPetersson,Mirzakhani:2006fta,Wolpert2010WeilPetersson,Eynard:2016yaa}.

This same formula~(\ref{eq:W11-E}) (or~(\ref{eq:W11-z})) works for various supersymmetric examples, perhaps surprisingly since they are governed by a different class of matrix model. For example for ${\cal N}{=}1$ JT supergravity~\cite{Mertens:2017mtv,Stanford:2017thb,Stanford:2019vob},  which was formulated in terms of a $u(x)$ in refs.~\cite{Johnson:2020heh,Johnson:2020exp}, the appropriate edge value is $\mu{=}1$, and while $u_0$ and its derivatives vanish at that point, meaning can be given to finite ratios of them (see Section~\ref{sec:N=1-super-cases}). It turns out that $u_0^{\prime\prime}/u_0^\prime{=}{-}3$ there, giving $V_{1,1}{=}{-}\frac18$, the known value. Finally, for ${\cal N}{=}{2}$ JT supergravity (formulated in terms of a function $u(x)$ in ref.~\cite{Johnson:2023ofr}), there are some more involved formulae (derived in ref.~\cite{Ahmed:2025lxe}, but not listed here)  for $u_0(x)$ and  $\mu$ which  yield: $u_0(\mu){=}E_0$, $u_0^\prime(\mu){=}{-}4\pi E_0$ and $u_0^{\prime\prime}(\mu){=}24\pi^2E_0{-}16\pi^4E_0^2$, whereupon  starting with (\ref{eq:W11-E}), defining $z^2{=}E_0{-}E$ and multiplying by Jacobian~${-}2z$, to give formula~(\ref{eq:W11-z}), yields (after dividing by an extra $2\pi$ to convert between conventions)   the  Laplace transform of:
\begin{equation}
\label{eq:V11-Neq2}
    V_{1,1}(b)=-\frac{1}{8}+\frac{(4\pi^2+b^2)}{48}E_0\ ,
    \quad\text{(${\cal N}{=}2$ JT/WP)}
\end{equation}
which was  derived by Turiaci and Witten~\cite{Turiaci:2023jfa} in studying the supergravity theory as a random matrix model.

There are many more examples that could be shown here, but these  will do for now. The point  is that  formula~(\ref{eq:W11-E}) is quite general and using the method of  ref.~\cite{Johnson:2024bue}, where it was derived, analogous expressions (for one boundary) for higher genus are readily obtained~\cite{Ahmed:2025lxe}, in a manner to be reviewed below.

Hopefully the  examples of the use of the formalism given above (and the many more in refs.\cite{Johnson:2024bue,Johnson:2024fkm,Lowenstein:2024gvz,Ahmed:2025lxe,Johnson:2026jbq,Johnson:2026plw}) strongly motivate the results of this paper: {\it Analogous formulae for all ${\widetilde W}_{g,n}$ can be readily derived.} 

Before explaining  how to construct them in general it is hard to resist showing some of the newly obtained  universal  formulae in action. Here  is the case of  genus one with two boundaries/energies, in terms of energies:
\begin{widetext}
\begin{eqnarray}
\label{eq:W12}
&&\hskip-1cm W_{1,2}(z_1,z_2)=
\frac{1}{z_1^2 z_2^2}
\left[
A_{12}
+ B_{12}\sum_{i=1}^2 \frac{1}{z_i^2}
+ C_{12}\sum_{i=1}^2 \frac{1}{z_i^4}
+ D_{12}\frac{1}{z_1^2 z_2^2}
\right]
\nonumber
\\
&&\hskip-1cm \text{with}\qquad
A_{12}= \frac{u_0^{(3)}(\mu)}{24\,u_0'(\mu)}-\frac{u_0''(\mu)^2}{24\,u_0'(\mu)^2}\ ,
\quad
B_{12}=-\frac{u_0''(\mu)}{8}\ ,
\quad
C_{12}=\frac{5}{32}u_0'(\mu)^2\ ,
\quad
D_{12}=\frac{3}{32}u_0'(\mu)^2\ ,
\end{eqnarray}
 and also in terms of boundary lengths:
\begin{eqnarray}
\label{eq:W12-ell}
&&\hskip-1cm{\widetilde W}_{1,2}(x,\ell_1,\ell_2)=
\frac{\sqrt{\ell_1\ell_2}}{24\pi}\,e^{-u_0(x)(\ell_1+\ell_2)}\;
\left[
\frac{u_0'''(x)}{u_0'(x)}
-\left(\frac{u_0''(x)}{u_0'(x)}\right)^2
-2u_0''(x)(\ell_1+\ell_2)
+\big(u_0'(x)\big)^2\big((\ell_1+\ell_2)^2-\ell_1\ell_2\big)
\right]\ .
\end{eqnarray}
\end{widetext}
In the first form~({\ref{eq:W12}}), $z_i^2{=}u_0(\mu){-}E_i$, and two Jacobian factors $(-2z_1)(-2z_2)$ were included to convert to the~$z_i$ dependent object. The formula for ${\widetilde W}_{1,2}(E_1,E_2)$ can readily be obtained by putting back those factors.  In the second form, the structure of the $\ell_i$ dependence generalizes that of the single boundary case~(\ref{eq:W11-ell}) rather interestingly. In both cases, the fact that there is an $\sqrt\frac{\ell}{\pi}$ overall factor for each boundary mimics the more familiar genus zero formula~(\ref{eq:W0n}). The reason for this will become clear later.

Let’s explore the new formula~(\ref{eq:W12}). Putting in the defining equation  ($u_0(x){=}{-}2x$) for the Airy model, four  terms vanish, leaving: 
\begin{eqnarray}
W_{1,2}^{\rm Airy}(z_1,z_2)&=&
\frac{1}{8\,z_1^2 z_2^2}
\left[
5\sum_{i=1}^2 \frac{1}{z_i^4}
+3\,\frac{1}{z_1^2 z_2^2}
\right]
\ ,
\end{eqnarray}
which is indeed the known result. It can be either considered as the genus one correction of a two point correlation function of the spectral density (after going back to the $(E_1,E_2)$ dependence, remembering the Jacobians,  appropriate continuation in sign of energy, and dividing by $\hbar^2$)  or it can be considered as a package containing these results from Kontsevich-Witten  intersection theory: $\langle \tau_2\tau_0\rangle_1
{=}
\langle \tau_1\tau_1\rangle_1
{=}
\langle \tau_0\tau_2\rangle_1
{=}
\frac{1}{24}
$. (This is because $5!!{=}15$ and $(3!)^2{=}9$ are the resulting combinatorial factors--see foonote~\ref{fn:combinatorics-note}--for the two types of term in the polynomial.)

To see all of the formula in action, let's insert the values~(\ref{eq:JT-Wp-u0-derivatives}) for JT gravity ({\it i.e.,} Weil-Petersson), giving:
\begin{eqnarray}
&&\hskip-0.5cm 
W_{1,2}^{\rm WP}(z_1,z_2)=\\
&&\hskip-0.1cm
\frac{1}{8\,z_1^2 z_2^2}
\left[
2\pi^4
+4\pi^2\sum_{i=1}^2 \frac{1}{z_i^2}
+5\sum_{i=1}^2 \frac{1}{z_i^4}
+3\,\frac{1}{z_1^2 z_2^2}
\right] 
\ .\nonumber
\end{eqnarray}
In analogy with the one-point example above, we are either computing a two point correction in the matrix model dual to JT gravity or (after double Laplace transform from $\{z_1,z_2\}$ to $\{b_1,b_2\}$) the Weil-Petersson volume $V_{1,2}^{\rm WP}$ of the moduli space $\overline{\cal M}_{1,2}$ of Riemann surfaces  with two geodesic boundaries at genus one:
\begin{equation}
    V^{\rm WP}_{1,2}(b_1,b_2)=
    \frac{\left(4 \pi^{2}+b_1^{2}+b_2^{2}\right) \left(12 \pi^{2}+b_1^{2}+b_2^{2}\right)}{192}\ .
\end{equation}
Formulae for more boundaries/energies at genus one are readily derived (we'll show how in the next section) but they grow very rapidly in size and so won't be shown here. The case $W_{1,3}$ is given in  Appendix~\ref{app:more-formulae}, with both energy and length dependent forms shown.\footnote{A perusal of the structure of the $W_{1,n}$ displayed so far will be rewarded with some patterns meeting the eye. The most obvious is that the lowest order coefficients $A_{1n}$ are simply related by being derivatives of each other. In fact $A_{1n}=\left(\frac{d}{dx}\right)^n\cdot F_1$ where $F_1{=-}\frac{1}{24}\ln u_0^\prime$, the genus one contribution to the free energy. More generally, $A_{gn}=\left(\frac{d}{dx}\right)^n\cdot F_g$, as will become clear in Section~\ref{sec:how-it-all-works}. Patterns of this kind will become stronger for the ${\cal N}{=}1$ supersymmetric cases in Section~\ref{sec:N=1-super-cases}, allowing for closed form (all-$n$) formulae to be written for a given $g$.}

The higher genus one-boundary cases $W_{2,1}$ and $W_{3,1}$ were first  worked out  in ref.~\cite{Ahmed:2025lxe}. They will be discussed in the next section, which will explain a more fundamental origin for them, and situate them all in a more complete framework.   To close this introduction, just in case the examples above were not enough, here is the new formula for the case of genus $g{=}2$ with two boundaries/energies:
\begin{widetext}

\begin{eqnarray}
\label{eq:W22-z-A}
&&\hskip-0.5cm W_{2,2}(z_1,z_2)
=-\frac{1}{z_1^2 z_2^2}\Bigg[
 A_{22}
+ B_{22}\sum_{i=1}^2\frac{1}{z_i^{2}}
+ C_{22}\sum_{i=1}^2\frac{1}{z_i^{4}}
+ D_{22}\frac{1}{z_1^2 z_2^2}
+ E_{22}\sum_{i=1}^2\frac{1}{z_i^{6}}
+ F_{22}\frac{1}{z_1^2 z_2^2}\sum_{i=1}^2\frac{1}{z_i^{2}}
\\
&&\hskip-0.1cm+ G_{22}\!\left(
231\sum_{i=1}^2\frac{1}{z_i^{8}}
+203\,\frac{1}{z_1^2 z_2^2}\sum_{i=1}^2\frac{1}{z_i^{4}}
+210\,\frac{1}{z_1^4 z_2^4}
\right)
+ H_{22}\!\left(
1155\sum_{i=1}^2\frac{1}{z_i^{10}}
+945\,\frac{1}{z_1^2 z_2^2}\sum_{i=1}^2\frac{1}{z_i^{6}}
+1015\,\frac{1}{z_1^4 z_2^4}\sum_{i=1}^2\frac{1}{z_i^{2}}
\right)
\Bigg] \ ,    \nonumber
\end{eqnarray}
where attention is drawn to the overall minus sign, and:
\begin{eqnarray}
\label{eq:W22-z-B}
&&A_{22}
=
\frac{1}{576}\,\frac{u_0^{(6)}(\mu)}{u_0'(\mu)^2}
-\frac{41}{2880}\,\frac{u_0''(\mu)u_0^{(5)}(\mu)}{u_0'(\mu)^3}
-\frac{73}{2880}\,\frac{u_0^{(3)}(\mu)u_0^{(4)}(\mu)}{u_0'(\mu)^3}
+\frac{17}{240}\,\frac{u_0''(\mu)^2u_0^{(4)}(\mu)}{u_0'(\mu)^4}
\nonumber
\\
&&\hspace{8.5cm}
+\frac{19}{192}\,\frac{u_0''(\mu)u_0^{(3)}(\mu)^2}{u_0'(\mu)^4}
-\frac{35}{144}\,\frac{u_0''(\mu)^3u_0^{(3)}(\mu)}{u_0'(\mu)^5}
+\frac{1}{9}\,\frac{u_0''(\mu)^5}{u_0'(\mu)^6},
\nonumber
\end{eqnarray}
\begin{eqnarray}
&&
B_{22}
=
-\frac{1}{96}\,\frac{u_0^{(5)}(\mu)}{u_0'(\mu)}
+\frac{1}{24}\,\frac{u_0''(\mu)u_0^{(4)}(\mu)}{u_0'(\mu)^2}
+\frac{11}{384}\,\frac{u_0^{(3)}(\mu)^2}{u_0'(\mu)^2}
-\frac{23}{192}\,\frac{u_0''(\mu)^2u_0^{(3)}(\mu)}{u_0'(\mu)^3}
+\frac{23}{384}\,\frac{u_0''(\mu)^4}{u_0'(\mu)^4},
\nonumber
\end{eqnarray}
\begin{eqnarray}
&&
C_{22}
=
\frac{11}{192}\,u_0^{(4)}(\mu)
-\frac{5}{96}\,\frac{u_0''(\mu)u_0^{(3)}(\mu)}{u_0'(\mu)}
+\frac{5}{192}\,\frac{u_0''(\mu)^3}{u_0'(\mu)^2}\ ,
\nonumber
\end{eqnarray}
\begin{eqnarray}
&&
D_{22}
=
\frac{3}{64}\,u_0^{(4)}(\mu)
-\frac{1}{64}\,\frac{u_0''(\mu)u_0^{(3)}(\mu)}{u_0'(\mu)},
\quad
E_{22}
=
-\frac{203}{768}\,u_0'(\mu)u_0^{(3)}(\mu)
-\frac{91}{768}\,u_0''(\mu)^2,
\nonumber\\
&&
F_{22}
=
-\frac{29}{128}\,u_0'(\mu)u_0^{(3)}(\mu)
-\frac{9}{64}\,u_0''(\mu)^2,
\qquad 
G_{22}=\frac{1}{256}\,u_0'(\mu)^2u_0''(\mu),
\qquad
H_{22}=-\frac{1}{2048}\,u_0'(\mu)^4.
\end{eqnarray}
In the Airy case, where everything involving derivatives higher than $u_0^\prime{=}{-}2$ vanishes,  we get the result:
\begin{eqnarray}
  W^{\rm Airy}_{2,2}&=&   
\frac{1}{128\,z_1^2 z_2^2}
\left[
1155\sum_{i=1}^2 \frac{1}{z_i^{10}}
+945\,\frac{1}{z_1^2 z_2^2}\sum_{i=1}^2 \frac{1}{z_i^6}
+1015\,\frac{1}{z_1^4 z_2^4}\sum_{i=1}^2 \frac{1}{z_i^2}
\right]
 \ ,
\end{eqnarray}
The JT gravity (Weil-Petersson) case provides all the extra non-zero derivatives, and inserting them  results in:
\begin{eqnarray}
 && W^{\rm WP}_{2,2}= 
 \frac{1}{z_1^2 z_2^2}
\Bigg[
\frac{787\pi^{10}}{480}
+\frac{1085\pi^{8}}{288}\sum_{i=1}^2 \frac{1}{z_i^2}
+\frac{551\pi^{6}}{72}\sum_{i=1}^2 \frac{1}{z_i^4}
+\frac{7\pi^{6}}{z_1^2 z_2^2}
+\frac{399\pi^{4}}{32}\sum_{i=1}^2 \frac{1}{z_i^6}
+\frac{181\pi^{4}}{16}\frac{1}{z_1^2 z_2^2}\sum_{i=1}^2 \frac{1}{z_i^2}
\\
&&\hspace{5cm}
+\frac{\pi^2}{16}\left(
231\sum_{i=1}^2 \frac{1}{z_i^8}
+203\,\frac{1}{z_1^2 z_2^2}\sum_{i=1}^2 \frac{1}{z_i^4}
+210\,\frac{1}{z_1^4 z_2^4}
\right)
\Bigg]
 +W_{2,2}^{\rm Airy}\ .
\end{eqnarray}

\end{widetext}
It is straightforward and amusing to yield such multi-boundary formulae for the ${\cal N}{=}2$ JT case as well, but we will refrain from doing so here, in order to save space.

Now that the motivation has been provided, the main task of the rest of the paper is to explain where all this structure comes from, and how to compute  general formulae for $W_{g,n}(\{ z_i\})$ (or equivalently ${\widetilde W}_{g,n}(\{E_i\})$), where $(i{=}1,\cdots,n)$. This is the purpose of the next section. 

\section{The Method}

It might be helpful  to gather together all in one place the key tools that are needed. Therefore  the next subsection is a swift review that the more expert  reader can skip, moving straight on to the derivation of the correlators $W_{g,n}$ in the following subsection.

\subsection{The basic toolbox}
\label{sec:toolbox}
Certain structures and powerful tools emerged in the early studies of random matrix models and the double scaling limit~\cite{Gross:1990vs,Gross:1990aw,Brezin:1990rb,Douglas:1990ve}. Key to all of it  was to formulate the matrix models in terms of a family of orthogonal polynomials~\cite{Meh2004}, $P_n(\lambda)$, where $\lambda$ is an eigenvalue, and the label $n{=}0,\ldots,\infty$. They are typically normalized so that $P_n(\lambda){=}\lambda^n+\text{lower powers}.$ The  matrix model's defining  probability (of a draw from the ensemble), often called the potential $V(\lambda)$, enters through the measure in the integral that expresses the polynomials' orthogonality.  The polynomials satisfy a recursion relation with two recursion coefficients that are determined by the matrix model potential:
\begin{equation}
    \lambda P_n(\lambda)=P_{n+1}(\lambda)+s_n P_n(\lambda)+r_nP_{n-1}(\lambda)\ .
\end{equation}Computations in the model hence require the  determination of the polynomials, which amounts to finding the recursion coefficients $(r_n,s_n)$.
The matrix model determines them through a difference equation whose structure is fixed by the potential $V(\lambda)$.

At large $N$, and in the double-scaling limit,  the discrete label $n$ indexing the polynomials becomes a continuous real parameter $x$, and the recursion coefficients (which also carry an orthogonal polynomial index) become functions of the $x$. Furthermore, in the double scaling limit, which amounts to focussing on the neighbourhood the edge of the leading spectral density (where universal physics can be found), a linear combination of the two recursion functions becomes the pertinent variable, and it is denoted as the function $u(x)$.\footnote{The other linear combination is relevant at the other end of the spectral density, now infinitely far away in this limit.}  It has a leading piece and  corrections:
\begin{equation}
\label{eq:u-corrections}
    u(x)=u_0(x)+\sum_{g=1}^\infty u_{2g}(x) \hbar^{2g}+\cdots\ ,
\end{equation}
where the ellipses denote non-perturbative parts that won't concern us here. The leading function $u_0(x)$ is (together with its derivatives)  what entered into all perturbative quantities in the previous section. 

Another  key structure is that there is an auxiliary Schr\"odinger Hamiltonian, where $u(x)$ is the potential:
\begin{equation}
\label{eq:aux-hamiltonian}
    {\cal H}=-\hbar^2\frac{\partial^2}{\partial x^2}+u(x)\ .
\end{equation} 

The matrix model itself constrains what $u(x)$ can be. The difference equation for the recursion coefficients become, in the limit,  a non-linear equation for $u(x)$ called  a ``string equation'',  of which the most well known example, from working with Hermitian random matrices, can be written as~\cite{Gross:1990vs,Gross:1990aw,Douglas:1990ve,Brezin:1990rb}:
\begin{equation}
    \label{eq:little-string-equation}
   {\cal R}\equiv \sum_{k=1}^\infty t_k R_{k}[u]+x=0\ ,
\end{equation} but another important example will appear below. Here $R_k[u]$ is the $k$th Gel'fand-Dikii polynomial in~$u(x)$ and its $x$-derivatives, normalized here so that the purely polynomial part is unity. The first few are~\cite{Gelfand:1975rn}:
 \begin{eqnarray}
 \label{eq:GD-polynomials}
 &&R_0[u]{=}1\ ,\quad R_1[u]{=}u\ ,\quad  R_2[u]{=}u^2{-}\frac{\hbar^2}{3}u^{\prime\prime}\ , \quad \text{and}\nonumber\\
 &&R_3[u]{=}u^3{-}\frac{\hbar^2}{2}(u^\prime)^2{-} {\tiny \hbar^2}uu^{\prime\prime}{+}\frac{\hbar^4}{10}u^{\prime\prime\prime\prime}\ .
 \end{eqnarray} A prime denotes an $x$-derivative, and each comes with an~$\hbar$. More generally, $R_k{=}u^k+\cdots\# \hbar^{2k-2}u^{(2k-2)}$, where $u^{(m)}$ means the $m$th derivative. The intermediate terms involve mixed orders of derivatives. The~$R_k[u]$ may be readily determined for any $k$ using a recursion relation, which is (in the normalization conventions of this paper):
 \begin{equation}
     \label{eq:recursion}
    R_{k+1}^{\prime} = \frac{2k + 2}{2k + 1} \left( \frac{1}{2} u^{\prime} R_k + u R_k^{\prime} - \frac{\hbar^2}{4} R_k^{\prime\prime\prime} \right)\ ,
 \end{equation} which can be used to derive a useful $\hbar^2$ expansion of the $R_k[u]$. The first few orders are~\cite{Johnson:2021owr}:\footnote{Note that ref.~\cite{Ahmed:2025lxe} explores the structure of higher orders and provides a useful algorithm for determining them.}
 \begin{equation}
\label{eq:Rk-expansion}
\begin{aligned}
{}&R_k[u]
=u^k
-\hbar^2\frac{k(k-1)}{12}\,
u^{k-3}\Bigl((k-2)(u')^2+2u\,u''\Bigr)
\\[1mm]
&\quad
+\hbar^4\frac{k(k-1)(k-2)}{1440}\,
u^{k-6}\Bigl[
24u^3u^{(4)}
+48(k-3)u^2u'u'''
\\[1mm]
&\quad
+36(k-3)u^2(u'')^2
+44(k-3)(k-4)u\,(u')^2u''
\\[1mm]
&\quad
+5(k-3)(k-4)(k-5)(u')^4
\Bigr]
+O(\hbar^6)\ ,
\end{aligned}
\end{equation}
which, for $k=0,1,2,3$ can be seen to truncate nicely to the expressions in~(\ref{eq:GD-polynomials}).

It is enough for our purposes to take the string equation shown above as defining $u(x)$ perturbatively in a large negative $x$ expansion about some leading solution defined by  ${\cal R}_0\equiv\sum t_k u_0^k+x=0$. What the particular~$t_k$ are depends upon the original matrix model potential. The Airy model example from before had all $t_k{=}0$ except $t_1{=}\frac12$. The Weil-Petersson/Jackiw-Teitelboim  example had~\cite{Dijkgraaf:2018vnm,Okuyama:2019xbv,Johnson:2019eik}
$t_k{=}\frac{\pi^{2k-2}}{2k!(k-1)!}$.

Another class of random matrix models  of interest here yields a different string equation~\cite{Morris:1990bw,Dalley:1991qg,Dalley:1992br}:
\begin{equation}
\label{eq:big-string-equation}
u{\cal R}^2-\frac{\hbar^2}2{\cal R}{\cal R}^{\prime\prime}+\frac{\hbar^2}4({\cal R}^\prime)^2=\hbar^2\Gamma^2\ , 
\end{equation}
where ${\cal R}$ is defined as in~(\ref{eq:little-string-equation}). The parameter~$\Gamma$ naturally has interpretations as either integer or half integer, and these models are in the $(\boldsymbol{\alpha},\boldsymbol{\beta}){=}(2\Gamma{+}1,2)$ case of the Altland-Zirnbauer classification~\cite{Altland:1997zz}. This  class of models can be understood as coming from generalizing Wishart-type models of {\it positive} random Hermitian matrices with~$\Gamma$ degenerate $E=0$ states. They are also equivalent to equations arising from random Unitary matrices with $2\Gamma{+}1$ quark flavours.~\cite{Myers:1991akt,Dalley:1992br,Gross:1991aj}

Equation~(\ref{eq:big-string-equation}) has   perturbative  expansions for $u(x)$ in either of the $x{=}\pm\infty$ regimes. All we will need here is that 
perturbation theory
for this equation has the same structure as that of  equation~(\ref{eq:little-string-equation}) in the $x\to-\infty$ regime, as well as (provided $u_0\neq0$) the $x\to+\infty$ regime.~\cite{Ahmed:2025lxe}\footnote{The case of perturbation theory about $u_0{=}0$, very important for {\it e.g.}, ${\cal N}{=}1$ supersymmetric cases, can also be treated, but is most efficiently done slightly differently from the $u_0{\neq}0$ cases. This is discussed in Section~
\ref{sec:N=1-super-cases}.}

Once a leading $u_0(x)$ is fixed, a string equation recursively defines the higher $u_{2g}(x)$ in the expansion~(\ref{eq:u-corrections}) in terms of $u_0(x)$ and its derivatives. Explicit expressions will be given later, but they are derived readily by expanding the string equation itself in powers of $\hbar^2$,  using~(\ref{eq:Rk-expansion}) for the $R_k$. 

Very importantly for what is to come, we should think of the framework as an ``off-shell'' system where a particular $\{t_k\}$
remains unspecified until the end. Hence,~$u(x)$ is not just a function of $x$, but also of the infinite set~$\{t_k\}$. 
Given that, we can ask how $u(x,\{t_k\})$  depends on a particular~$t_k$. The answer is that it changes according to the KdV heirarchy of flows:
\begin{equation}
    \label{eq:kdv-flows}
    \frac{\partial u}{\partial t_k}=R_{k+1}^\prime[u]\ ,
\end{equation}
and given that $R_1{=}u$, we can see that $x\equiv t_0$, although in many formulae the explicit $x$ dependence will still be indicated alongside dependence on the set $\{t_k\}$.

The function $u(x)$ is itself the second derivative of the generating function of connected closed surfaces, {\it i.e.,} the free energy~$F$:
\begin{equation}
\label{eq:free-energy}
    u(x)=2\hbar^2\frac{\partial^2 F}{\partial x^2}\ ,
\end{equation}
\medskip
\noindent where the $t_k$ dependence is implicit on both sides. By virtue of $u(x)$'s expansion~(\ref{eq:u-corrections}), the free energy $F$ also has a genus expansion:
\begin{equation}
\label{eq:uF-corrections}
    F=\sum_{g=0}^\infty F_{g}\hbar^{2g-2}+\cdots\ ,
\end{equation}
and explicit expressions for the $F_g$ in terms of $u_0$ and its derivatives will appear later. (They will follow from the fact that the $u_{2g}(x)$ can be written in this way.)

Derivatives of $F$ with respect to a $t_k$ can be thought of as dual to inserting a point-like operator ${\cal O}_k$ in the 2D gravity theory. 
Those point-like operators are sometimes called ``microscopic loops'' (which can be traced back to the 't Hooftian diagrammatics). Our main interest will be loops of finite size in the double scaling limit, called ``macroscopic loops''\cite{Banks:1990df}. It is natural to ask about the expectation value of a loop of length $\ell$, denoted ${\widetilde W}(\ell)$ in this framework. An equivalent formulation is in terms of the object ${\widetilde W}(E)$ obtained by Laplace transforming  from $\ell$ to~$E$. Up to a continuation from $E\to-E$ this is essentially the spectral density.

In general, the correlator of $n$ such loops is:
\begin{equation}
    \label{eq:loop-correlator}
   {\widetilde W}(E_1\cdots E_i)\equiv\hbar^2(-1)^{n-1}\delta_{E_1}\cdots\delta_{E_n} F(x,\{ t_k\})\Big|_{\mu}\ ,
\end{equation}
(the sign factors are present to match standard conventions in the literature) and the loop operator $\delta_{E_i}$ will be a key workhorse in this paper. In the conventions of this paper it is~\cite{Banks:1990df,Dijkgraaf:1991rs}:
\begin{equation}
    \label{eq:loop-operator}
     \delta_{E_i}=\sum_{k=0}^\infty \frac{2C_{k+1}}{(-E_i)^{k+\frac32}}\frac{\partial}{\partial t_k}\ ,\,\,\text{with}\,\,\, C_k=(-1)^k\frac{(2k-1)!!}{2^{k+1}k!}\ .
\end{equation}

All the tools needed to define the method for deriving the formulae seen in the previous section (and in the Appendices) are now in place.

\begin{widetext}
\subsection{How it all works}
\label{sec:how-it-all-works}
\subsubsection{One-boundary review}
Let's  start by inserting one loop. This will quickly show why the Gel'fand-Dikii equation central to ref.~\cite{Johnson:2024bue}'s method is relevant. We first get:

\begin{eqnarray}
    \label{eq:loop-correlator}
   {\widetilde W}(x,E_1)\equiv\hbar^2\delta_{E_1}F
   &=&\int^x \!dx' \!\int^{x'}\! dx^{\prime\prime}\left(
   \sum_{k=0}^\infty \frac{C_{k+1}}{(-E_1)^{k+\frac32}}\frac{\partial u(x^{\prime\prime})}{\partial t_k}\right)
   =\int^x \!dx' \!\int\! dx^{\prime\prime}\left(
   \sum_{k=0}^\infty \frac{C_{k+1}}{(-E_1)^{k+\frac32}}\frac{\partial R_{k+1}[u]}{\partial x^{\prime\prime}}\right)
   \nonumber \\ 
   &=&\int^x \!dx' \!\left(
   \sum_{k=0}^\infty \frac{C_{k}R_{k}[u(x')]}{(-E_1)^{k+\frac12}}\right) 
   =\int^x  {\widehat R}(x',E_1) dx'  
   \ ,
\end{eqnarray}
\end{widetext}
where in the first and second steps equations~(\ref{eq:free-energy}) and~(\ref{eq:kdv-flows})  were used, the next line used that $R_0{=}1$, and the last line is the known expansion  of ${\widehat R}(x,E_1)$,  the diagonal resolvent of Hamiltonian~(\ref{eq:aux-hamiltonian}). That expansion was introduced by Gel'fand and Dikii in ref.~\cite{Gelfand:1975rn}. They used  a different normalization of the $R_k[u]$, but the numbers~$C_k$ built into our definition of the operator~(\ref{eq:loop-operator}) precisely restores that normalization.

The ordinary differential equation (ODE) that ${\widehat R}(x,E_1)$ obeys is~\cite{Gelfand:1975rn}:
\begin{equation}
\label{eq:gelfand-dikii}
    4(u-E_1)\widehat{R}^2-2\hbar^2\widehat{R}\widehat{R}^{\prime\prime}+\hbar^2(\widehat{R}^\prime)^2=1\ ,
\end{equation}
and is sometimes called the Gel'fand-Dikii equation. 
It takes as input a given potential $u(x)$, and is  useful in a range of physics applications.

We will use it to carefully organize (genus) perturbation theory here, following ref.~\cite{Johnson:2024bue}, and we will see further that it will also yield a simple treatment of the loop operator perturbatively too. Recall that $u(x)$  has a perturbative expansion in powers of~$\hbar$.  Moreover, at every order in the expansion of ${\widehat R}(x,E_1)$  in powers of $E_1$, there are additional powers of $\hbar$ coming from the structure of the $R_k[u]$ (see equation~(\ref{eq:GD-polynomials})). The key idea  is that the  organization that is interesting to us is the $\hbar$ (genus) expansion and so it makes sense to entirely re-expand  ${\widehat R}(x,E_1)$ to make it manifest, defining:  
 \begin{equation}
 \label{eq:R-expansion}
     \widehat{R}(x,E_1)=\sum_{g=0}^\infty \hbar^{2g}\widehat{R}_g(x,E_1)+\cdots\ ,
 \end{equation}
(the ellipsis denotes non-perturbative parts).\footnote{Recently, ref.~\cite{Johnson:2026jbq} explored and extracted useful non-perturbative physics from the Gel'fand-Dikii equation in this context, naturally extracting (for example) formulae for the large $g$ growth of  several different types of WP volume.}  
The leading terms in the expansion for ${\widehat R}_g(x,E)$ may be efficiently developed by simply recursively solving equation~(\ref{eq:gelfand-dikii}):
\begin{widetext}  
\begin{eqnarray}
\label{eq:useful-expansion}
&& {\widehat R}(x,E_1)
= -\frac12\frac{1}{[u_0(x)-E_1]^{1/2}}
+
\frac{\hbar^2}{64}\left\{
\frac{16u_2(x)}{[u_0(x)-E_1]^{3/2}}
+\frac{4u^{\prime\prime}_0(x)}{[u_0(x)-E_1]^{5/2}}-\frac{5(u^{\prime}_0(x))^2}{[u_0(x)-E_1]^{7/2}}\right\}
\\
&&\hskip+2.5cm
+\frac{\hbar^4}{4096}\left\{
\frac{1024u_4(x)}{[u_0(x)-E_1]^{3/2}}
-\frac{256[3u_2(x)^2-u^{\prime\prime}_2(x)]}{[u_0(x)-E_1]^{5/2}}
+\frac{64[u^{(4)}_0(x) -10u_2(x) u^{''}_0(x)-10u^\prime_0(x)u^\prime_2(x)]}{[u_0(x)-E_1]^{7/2}}\right. \nonumber \\
&&\hskip+2.5cm
\left.
-\frac{16[28u^{(3)}_0(x)u^\prime_0(x)+21u^{\prime\prime}_0(x)^2-70u_2(x)u_0(x)^2]}{[u_0(x)-E_1]^{9/2}}
+\frac{{1848}u^{\prime}_0(x)^2u^{\prime\prime}_0(x)}{[u_0(x)-E_1]^{11/2}}-\frac{1155u^{\prime}_0(x)^4}{[u_0(x)-E_1]^{13/2}}\right\} +\cdots\ .   \nonumber
\end{eqnarray} 
This looks like an unhelpful mess, but  a key piece of information  has not been used yet. We expect that $u(x)$ satisfies a string equation, at least perturbatively. That equation  determines recursively the higher $u_{2g}(x)$ in terms of the leading solution $u_0(x)$. For example:
\begin{equation}
\label{eq:eliminate-u2-u4}
    u_2(x)=\frac{u_0''^2-u_0'u_0'''}{12u_0'^2}=-\frac{1}{12}\frac{d^2}{dx^2}\ln(u_0'(x))\ ,\quad
\text{and}\quad 
    u_4(x)=\frac{d^2}{dx^2}\left[\frac{u_0''(x)^3}{90 u_0'(x)^4}-\frac{7 u_0^{(3)}(x)
   u_0''(x)}{480 u_0'(x)^3}+\frac{u_0^{(4)}(x)}{288
   u_0'(x)^2}\right]\ .
\end{equation}
Remarkably, once those are substituted into~(\ref{eq:useful-expansion}), all the ${\widehat R}_g(x,E_i)$ for $g>1$ become total derivatives!~\cite{Johnson:2024bue}
Specifically,
         \begin{equation}
         {\widehat R}_g(x,E_1) = \frac{d}{dx}\left[{\widehat Q}_g(x)\right]\ ,
         \label{eq:no-longer-a-total-surprise}
     \end{equation}
where:
\begin{eqnarray}
\label{eq:totally-awesome}
&&\hskip-1.1cm 
   \widehat{Q}_1(x)=-\frac{u_0^{\prime\prime}(x)}{48 u_0^\prime(x)[u_0(x)-E_1]^{3/2}}+\frac{u_0^\prime(x)}{32[u_0(x)-E_1]^{5/2}} \ ,
\\
   &&\hskip-1.1cm 
   \widehat{Q}_2(x)=\frac{105u_0'^3}{2048[u_0-E_1]^{11/2}}-\frac{203u_0'u_0''}{3072[u_0-E_1]^{9/2}}+\frac{29u_0'u_0'''-3u_0''^2}{1536u_0'[u_0-E_1]^{7/2}} \\
        &&\hskip-1.0cm-\frac{17u_0''^3-34u_0'u_0''u_0'''+15u_0'^2u_0^{(4)}}{3840u_0'^3[u_0-E_1]^{5/2}}-\frac{64u_0''^4-111u_0'u_0''^2u_0'''+21u_0'^2u_0'''^2+31u_0'^2u_0''u_0^{(4)}-5u_0'^3u_0^{(5)}}{5760u_0'^5[u_0-E_1]^{3/2}}\ .
        \nonumber
\label{eq:totally-awesome-two}
\end{eqnarray}
\end{widetext}
and ${\widehat Q}_3(x)$ is given explicitly in Appendix~\ref{app:more-formulae}. 
While it had not been proven that this total derivative phenomenon happens to all orders (we will change that situation shortly), it is consistent with various anticipated features of the results.
In particular, it is clear that since the ${\widehat R}_g(x,E_1)$ are total derivatives, the last integral in equation~(\ref{eq:loop-correlator}) can be performed for each $g{>}0$ order, giving only a boundary term that is polynomial in inverse powers of $z_1{=}(u_0(\mu){-}E_1))^\frac12$. As noted in ref.~\cite{Johnson:2024bue} this is consistent with the expectation that the $W_{g, 1}(z_1)$ are polynomials of that form (which for the Weil-Petersson case in turn follows from the result that the volumes are polynomial in the boundary lengths). 
More generally we might hope that a general formula for $W_{g,n}(\{z_i\})$  is also polynomial in inverse powers of all the $z_i{=}(u_0(x){-}E_i)^\frac12$. This will turn out to be true, as it must.

So finally then, we can define, order by order in perturbation theory, the ${\widetilde W}_{g,1}(E_1)$ by evaluating the $Q_g(x)$ at $x{=}\mu$, giving general formulae for the one-boundary correlators. Indeed the paper began by exhibiting 
${\widetilde W}_{1,1}(E_1)$, in equation~(\ref{eq:W11-E}), which comes from~(\ref{eq:totally-awesome}). We shall see later that the $\widehat{Q}_g$ themselves, although  they are key structures that arise from solving the Gel'fand-Dikii equation, are not fundamental and are built from simpler objects. 

Notice that  genus zero is special. The first term in (\ref{eq:useful-expansion}) is not a total derivative, but this is as it should be since putting it into (\ref{eq:loop-correlator}) gives precisely (after continuing~$E{\to}{-}E$ and dividing by $\pi\hbar$) the spectral density~(\ref{eq:integral-representation}).

\subsubsection{Multiple boundaries}
Now it is time to move swiftly on to adding more boundaries, this paper's focus. What is needed is simple to state: Just act further with the loop  operator~(\ref{eq:loop-operator}) as many times as needed, remembering (crucially) to allow $u(x,\{t_k\})$ its freedom to vary with respect to its arguments, only picking $x{=}\mu$  and the set $\{t_k\}$ at the end. 

%
In principle, one must determine how $\delta_{E_i}$ acts on ${\widetilde W}_1(x,u(x),E_1){=}\int^x {\widehat R}(x^\prime,E_1) dx^\prime $. One could work out an all-orders expression for that and then expand the resulting object in $\hbar$ along the lines done for one boundary, finally imposing the fact that each appearance of $u_{2g}(x)$ be replaced with expressions involving only $u_0(x)$ and its derivatives.  This in principle could work. However, this step was organizationally efficient for the one boundary insertion previously because the Gel'fand-Dikii ODE made producing the expansion in powers of $\hbar$ particularly easy. There's not obviously a nice object solving some ODE to work with here.

Another option is to work order by order, acting with the loop operator on the already $\hbar$-expanded expression. One then has to work out the action of $\delta_{E_i}$ order by order on the various $u_{2g}$, which presumably mixes with other~$u_{2g'}$, finally then using the string equation condition at the end. 
The good news is that the Gel'fand-Dikii resolvent (and the equation it satisfies) already provides the efficient organization of this approach.

The first step to see this is to  revisit equation~(\ref{eq:loop-correlator}) and take two $x$-derivatives,  giving:
\begin{eqnarray}
    \delta_{E_i}(u(x))&=&2\frac{d}{dx}{\widehat R}(x,E_i) \nonumber\\
    &=& \frac{1}{2}\frac{u^\prime_0(x)}{(u_0(x)-E_i)^\frac32} +2\sum_{g=1}^\infty\hbar^{2g}\frac{d}{dx}{\widehat R}_g(x,E_i)\ .
\end{eqnarray}
Now that we've applied the loop operator we are free again to use the string equation to eliminate all higher~$u_{2g}(x)$ in terms of $u_0(x)$, and  
decomposing $u(x){=}\sum_g u_{2g}(x)\hbar^{2g}$ shows that:
\begin{eqnarray}
   &&\hskip-0.75cm \delta_{E_i}(u_0(x))=\frac{1}{2}\frac{u^\prime_0(x)}{(u_0(x)-E_i)^\frac32}\ ,\quad\text{and}
    \label{eq:delta-u0}\\
    &&\hskip-0.75cm \delta_{E_i}(u_{2g}(x))=2\frac{d}{dx}{\widehat R}_g(x,E_i)=
2\frac{d^2}{dx^2}{\widehat Q}_g(x,E_i)\ .
    \label{eq:delta-u2g}
\end{eqnarray}
This tells us something rather simple. After forming the one-boundary objects, and having eliminated all~$u_{2g}(x)$ using the string equation in favour of  $u_0(x)$ and its derivatives, the action $\delta_{E_i}(u_0)$ in~(\ref{eq:delta-u0}) is all that we need! It is also pleasing that it is an exact simple form that is  guaranteed to preserve the fact that our loop objects will be polynomial in inverse powers of the $z_i{=}(u_0(x){-}E_i)^{\frac12}$. In fact, a direct way of seeing the action on $u_0(x)$ is to take the leading piece of the KdV flows at the outset, writing:
\begin{eqnarray}
\hskip-0.4cm \delta_{E_i}(u_0(x))&=&\sum_{k=0}^\infty \frac{2C_{k+1}}{(-E_i)^{k+\frac32}}\frac{\partial u_0}{\partial t_k}
 \nonumber\\
 &=&\sum_{k=0}^\infty \frac{2C_{k+1}}{(-E_i)^{k+\frac32}}(k+1)u_0^ku_0^\prime\nonumber\\
 &=&\frac{u_0^\prime}{(-E_i)^\frac32}\sum_{k=0}^\infty2(k+1)C_{k+1} \left(\frac{u_0}{-E_i}\right)^k\nonumber\\
 &=&\frac12\frac{u_0^\prime}{(-E_i)^\frac32}
 \left(1-\frac{u_0}{E_i}\right)^{-\frac32}\nonumber\\
 &=&\frac12\frac{u_0^\prime(x)}{(u_0(x)-E_i)^\frac32}\ .
 \label{eq:loop-action-u0}
\end{eqnarray}
In the second line the KdV flows were used, along with the leading form for $R_{k+1}[u_0]=u_0^{k+1}+\cdots$. To get from the third to the fourth, notice that given the form of~$C_k$ given in equation~(\ref{eq:loop-correlator}), $4(k+1)C_{k+1}$ turns out to be precisely the binomial coefficient for fractional power $-\frac32$.

There is an important class of  models where $u_0(x)$ is exactly zero. Such models can still be treated in this framework, either as a limiting case, or more directly by simply working with higher non-vanishing $u_{2g}$ and the relations between them,  and using action~(\ref{eq:delta-u2g}) whenever they appear.  Moreover, the relative simplicity of such models allows some powerful closed form (all-$n$) formulae  to be readily derived. See Section~\ref{sec:N=1-super-cases} for discussion of such cases.

So finally the path is clear. The Gel'fand-Dikii equation does the heavy lifting for giving the genus-by-genus proto-one-boundary objects ${\widetilde W}_{g,1}(x,E_1){=}{\widehat{Q}}_g(x,E_1)$, built from $u_0(x)$ and its derivatives. Evaluating them at $x{=}\mu$ gives the one-boundary correlators.  Before evaluating them however, we can act on them $n$ times, using the simple action~(\ref{eq:delta-u0}) of the boundary operator on $u_0(x)$, and the fact that it commutes with derivatives, to get, for $g{>}0$:
\begin{equation}
    {\widetilde W}_{g,n}(\{E_i\})=(-1)^{n-1}\delta^{(u_0)}_{E_n}\cdots\delta^{(u_0)}_{E_2} \cdot {\widehat{Q}}_g(x,E_1)\Big|_{x=\mu}\ ,
 \end{equation}
where $\delta^{(u_0)}_{E_i}$ acting on a function $f(u_0)$ means to act on~$u_0$ within that function using~(\ref{eq:delta-u0}).
Staring at this equation for a moment provokes the following thought. It would be nicer and more symmetric if the ${\widehat Q}_g$ themselves came from something more fundamental, also built entirely out of $u_0(x)$, that simply gets acted on by $\delta^{(u_0)}_{E_1}$. Some reflection reveals what it must be. It the free energy itself, but (of course!) written in terms of $u_0(x)$ and its derivatives by using the string equation. 
%
Looking back at equations~(\ref{eq:eliminate-u2-u4}) makes it all manifest. In both cases, the equation  for $u_{2g}(x)$ in terms of $u_0(x)$ and its derivatives is a second derivative of nothing other than~$2F_g$ ({\it c.f.} equation~(\ref{eq:free-energy})):
\begin{equation}
\label{eq:u2g-F}
    u_{2g}(x)=2\frac{d^2}{dx^2}F_{g}[u_0(x),u_0^\prime...]\ .
\end{equation}
So then acting on $F_g$ with the loop operator is just acting with our simple loop operator $\delta_{E_1}(u_0)$ through $F_g$'s $u_0(x)$ dependence. Let's test this.

\begin{widetext}
\noindent 
We see from~(\ref{eq:eliminate-u2-u4}) that  $F_1{=}{-}\frac{1}{24}\ln(u_0^\prime(x))$, and hence:
\begin{equation}
    {\widetilde W}_{1,1}(x,E_1)=\delta_{E_1}^{(u_0)}\cdot F_1[u_0^\prime]=-\frac{1}{24}\cdot\frac12\frac{1}{u_0^\prime}\frac{d}{dx}\left(\frac{u_0^\prime}{(u_0-E_1)^\frac32}\right)\ ,
\end{equation}
which gives precisely ${\widetilde Q}_1$ of equation~(\ref{eq:totally-awesome}), and so we recover~(\ref{eq:W11-E}) again. 

\noindent The same exercise works after writing $u_{4}(x){=}2\partial_x^2 F_2[u^{\prime}_0,u^{\prime\prime}_0,u^{(3)}_0,u^{(4)}_0]$
where:
\begin{equation}
    F_2[u^{\prime}_0,u^{\prime\prime}_0,u^{(3)}_0,u^{(4)}_0]=
\frac{(u_0'')^3}{180\,u_0'^4}
-\frac{7\,u_0^{(3)}u_0''}{960\,u_0'^3}
+\frac{u_0^{(4)}}{576\,u_0'^2}\ ,
\end{equation}
and there is the result $\delta_{E_1}^{(u_0)}\cdot F_2={\widehat Q}_2$ given in~(\ref{eq:totally-awesome-two}). 
Indeed, in the light of this, the  very long expression for the ${\widehat Q}_3$ listed in Appendix~\ref{app:more-formulae} (orignally derived in ref.~\cite{Ahmed:2025lxe}) can be spectacularly shortened to:
\begin{eqnarray}
    &&{\widehat Q}_3(x)=\delta_{E_1}\cdot F_3[u_0^\prime,u_0^{\prime\prime},\cdots]\ , \quad\text{where:}
    \nonumber\\\nonumber
   &&
   F_3[u_0,u_0^\prime,\cdots]=
-\frac{u_0^{(7)}}{10368\,u_0'^3}
+\frac{7\,u_0^{(6)}u_0''}{5760\,u_0'^4}
+\frac{53\,u_0^{(5)}u_0^{(3)}}{20160\,u_0'^4}
-\frac{353\,u_0^{(5)}u_0''^2}{40320\,u_0'^5}
+\frac{103\,(u_0^{(4)})^2}{60480\,u_0'^4}
-\frac{1273\,u_0^{(4)}u_0^{(3)}u_0''}{40320\,u_0'^5}
+\frac{83\,u_0^{(4)}u_0''^3}{1890\,u_0'^6}
\nonumber\\
&&\hskip0.5cm\quad
-\frac{59\,(u_0^{(3)})^3}{8064\,u_0'^5}
+\frac{83\,(u_0^{(3)})^2u_0''^2}{896\,u_0'^6}
-\frac{59\,u_0^{(3)}u_0''^4}{378\,u_0'^7}
+\frac{5\,u_0''^6}{81\,u_0'^8}\ .
\end{eqnarray}
Knowing this explicit form for $F_3$  will allow {\it all} genus three correlators to be computed. For similar future use at genus 4,  $F_4$ is listed as equation~(\ref{eq:F4}) in Appendix~\ref{app:free-energy-corrections}.

\end{widetext}

So now we can restate our formula for all correlators more symmetrically:
\begin{equation}
\label{eq:Wgn-E}
    {\widetilde W}_{g,n}(\{E_i\})=(-1)^{n-1}\delta^{(u_0)}_{E_n}\cdots\delta^{(u_0)}_{E_1} \cdot F_g[u_0^{\prime},u_0^{\prime\prime},\cdots]\Big|_{x=\mu}\ ,
\end{equation}
where $\delta^{(u_0)}_{E_i}$ is the very simple operator~(\ref{eq:delta-u0}). The action of $\delta^{(u_0)}_{E_i}$ will produce denominators which are powers of $z_i{=}(u_0(\mu){-}E_i)^\frac12$ (the number of powers increasing with~$g$ and~$n$).
Finally, if desired, we obtain the quantities  that are presented in terms of the $z_i$ (the $W$s without tildes, more natural in the topological recursion language~\cite{Eynard:2007kz}) by changing variables and multiplying by the Jacobian factors:
\begin{equation}
  W_{g,n}(z_i,\cdots, z_n)=
 \prod_{i=1}^n(-2z_i)\times{\widetilde W}_{g,n}(E_i,\cdots E_n)\ .
\end{equation}
This was how the general formulae~(\ref{eq:W12}),~(\ref{eq:W22-z-A}), and~(\ref{eq:W22-z-B}) were derived, as well as the additional examples in  Appendix~\ref{app:more-formulae}.
Notice that since they all commute (by virtue of their origin as sums of derivatives), the action of the $\delta^{(u_0)}_{E_i}$ guarantees that the $W_{g,n}$ are symmetric polynomials in inverse powers of $z_i$. (This is equivalent, in the JT/WP case, to Mirzakhani's discovery~\cite{Mirzakhani:2006fta} that the volumes are symmetric polynomials in the geodesic boundary lengths~$b_i$.)

Let's clearly note here that  we have cleared up the origin of the apparent miracle that the Gel'fand-Dikii resolvent becomes  a total derivative~(\ref{eq:no-longer-a-total-surprise}). We've shown that it   is indeed true to all orders, and moreover, that it was to be expected. It simply follows from the fact that the free energy $F{=}\sum_{g=0}\hbar^{2g-2}F_g$ can be written, order by order, as purely a function of $u_0(x)$'s derivatives, which in turn follows from the string equation. This then descends, after one action by the loop operator, to a statement about ${\widehat R}(x,E_1)$ to all orders, since the one-boundary loop is indeed the integral of ${\widehat R}(x,E_1)$ (see~(\ref{eq:loop-correlator})). Since we've shown that $\delta^{(u_0)}_{E_1}F_g{=}{\widehat Q}_g(x,E_1)$, the only way purely~$Q_g$ terms appear as the {\it integral} of each ${\widehat R}_g$ is if ${\widehat R}_g{=}\frac{d{\widehat Q}_g}{dx}$.

Put differently, inserting that $u_{2g}(x)$ satisfies the string equation can either be done at the level of $F_g$, and then one acts on it with the boundary operator $\delta_{E_i}$ to get ${\widehat Q}_g$, or one acts on  $F_g$ first with  $\delta_{E_i}$, getting $\int {\widehat R}_g(x,E_1)dx$, and then the requirement that $u(x)$ satisfies the string equation results in ${\widehat R}_g{=}\frac{d{\widehat Q}_g}{dx}$.

\begin{widetext}

\subsubsection{Multiple boundaries, genus zero}
We should handle the genus zero case, which is slightly different since it is an integral. We found that the the $\hbar^0$ part of ${\widetilde W}(x,E_1)$ was  just 
\begin{equation}
    {\widetilde W}_{0,1}(x,E_1)=-\frac12\int^x\frac{dx'}{\sqrt{u_0(x')-E_1}}\ ,
\end{equation} and so acting with $-\delta^{(u_0)}_{E_2}$ will define (according to (\ref{eq:Wgn-E}) and (\ref{eq:Wgn-E}))   ${\widetilde W}_{0,2}$. This cylinder case is special however (it is an ``unstable'' diagram in that $2g-2+n$ is not above zero), and to match standard conventions we will take the opposite sign, defining ${\overline W}_{0,2}={-}{\widetilde W}_{0,2}$:

\begin{eqnarray}\label{eq:two-point}
&&
{\overline W}_{0,2}(x,E_1,E_2)=
\frac18\int^{x}\!\!\frac{u_0'(x')\,dx'}{(u_0(x')-E_1)^\frac32(u_0(x')-E_2)^\frac32}
\nonumber\\&&\hskip2.5cm
=\frac18\int^{u_0(x)}
\!\!\frac{du_0}{(u_0-E_1)^\frac32(u_0-E_2)^\frac32}
=\frac{2(E_1+E_2-2u_0(x))}{8(E_1-E_2)^2\,\sqrt{u_0(x)-E_1}\,\sqrt{u_0(x)-E_2}} \ .
\end{eqnarray} Substituting either the Airy  or JT case gives the well-known universal result~\cite{Brezin:1993qg,BEENAKKER1,BEENAKKER2,BEENAKKER3} for this class of matrix models:
\begin{equation}
  {\overline W}_{0,2}(E_1,E_2)=  \frac{E_{2}+E_{1}}{4 \sqrt{-E_{1}}\, \sqrt{-E_{2}}\, \left(E_{1}-E_{2}\right)^{2}}\ .
\end{equation}
Henceforth all subsequent correlators will again be just in terms of inverse powers of $(u_0(x)-E_i)^\frac12$ and $u_0(x)$'s  derivatives. No integrals need trouble us any more.
For example, acting one more time with $\delta_{E_3}$, some nice cancellations in the   algebra give a pleasant result for the fundamental ``pair of pants'' diagram:
\begin{eqnarray}
   {\widetilde W}_{0,3}(x,E_1,E_2,E_3)
   =\hskip0.0cm\frac{u_0^{\prime}(x)}{16\,(u_0(x)-E_1)^{3/2}(u_0(x)-E_2)^{3/2}(u_0(x)-E_3)^{3/2}}\ , 
\end{eqnarray}
leading to:
\begin{eqnarray}
   { W}_{0,3}(z_1,z_2,z_3)
   =-\frac{u_0^\prime(x)}{2}\frac{1}{z_1^2z_2^2z_3^2}\ , \quad\text{which gives}\quad
{ W}_{0,3}(z_1,z_2,z_3)=\frac{1}{z_1^2z_2^2z_3^2}\ ,\end{eqnarray}
for Airy and JT, giving the standard $V_{0,3}{=}1$ normalization there. 
Another action with $-\delta_{E_4}$ gives, after some algebra:
\begin{eqnarray}
&&\hskip-0.85cm{\widehat W}_{0,4}(x,E_1,E_2,E_3,E_4)=
-\frac{1}{32\prod_{i=1}^4\big(u_0(x)-E_i\big)^{5/2}} \Bigg[
\prod_{i=1}^4\big(u_0(x)-E_i\big)\,u_0''(x)
-\frac{3}{2}u_0'(x)\frac{d}{dx}\Biggl(\prod_{i=1}^4\big(u_0(x)-E_i\big)\Biggr)
\Bigg]\ .
\end{eqnarray}
After a bit of thought, it becomes clear that it is instructive (and shorter) if these are written (using Laplace transforms) in terms of length variables as:
\begin{eqnarray}
\label{eq:low-W0nE}
&&\overline{W}_{0,2}(\ell_1,\ell_2)=     \frac{\sqrt{\ell_1\ell_2}}{2\pi(\ell_1+\ell_2)}
{\rm e}^{-u_0(x)(\ell_1+\ell_2)}\ ,\qquad
 W_{0,3}(\ell_1,\ell_2,\ell_3)= -    \frac{\sqrt{\ell_1\ell_2\ell_3}}{2\pi^{3/2}}\,
u_0'(x)\,
e^{-u_0(x)(\ell_1+\ell_2+\ell_3)}
\ ,\\
&& W_{0,4}(\ell_1,\ell_2,\ell_3,\ell_4)=   \frac{\sqrt{\ell_1\ell_2\ell_3\ell_4}}{2\pi^2}\,
e^{-u_0(x)(\ell_1+\ell_2+\ell_3+\ell_4)}
\Big[
-u_0''(x)
+\big(u_0'(x)\big)^2(\ell_1+\ell_2+\ell_3+\ell_4)
\Big]\ .
\end{eqnarray}
and the pattern here and henceforth is that summarized in equation~(\ref{eq:W0n}).
This suggests that we should seek a form of the loop operator that refers directly to $\ell_i$ variables. 

\end{widetext}

\subsubsection{Loop operator in length variables} The desired operator, $\delta^{(u_0)}_{\ell_i}$ would act on $u_0(x)$ and return a new function of $u_0(x)$ and $\ell_i$, just as its cousin~$\delta^{(u_0)}_{E_i}$ did (but with $u_0(x)$   and energy). Simply Laplace transforming does the trick:
\begin{equation}
    \label{eq:delta-L}
    \delta^{(u_0)}_{\ell_i}(u_0(x))={\mathcal L}\Bigl\{\delta^{(u_0)}_{E_i}(u_0(x))\Bigr\} = \sqrt{\frac{\ell_i}{\pi}}\partial_x{\rm e}^{-\ell_iu_0(x)}\ ,
\end{equation}
remarkably simple and exact. Like $\delta^{(u_0)}_{E_i}(u_0(x))$ it can be used with the chain rule to act on any function of $u_0(x)$ and its derivatives (as it commutes with differentiation). Hence, we can write a counterpart to equation~(\ref{eq:Wgn-E}):
\begin{equation}
\label{eq:Wgn-ell}
    {\widetilde W}_{g,n}(\{\ell_i\})=(-1)^{n-1}\delta^{(u_0)}_{\ell_n}\cdots\delta^{(u_0)}_{\ell_1} \cdot F_g[u^\prime_0,u_0^{\prime\prime},\cdots]\Big|_{x=\mu}\ .
\end{equation}
The genus zero version of this (for $n\geq2$) readily gives the famous formula~(\ref{eq:W0n}), as can be seen by studying the successive action of $\delta^{(u_0)}_{\ell_i}$ on $W_{0,2}$ in equation~(\ref{eq:low-W0nE}).

Intriguingly then, it can be applied anywhere in the chain of $W_{g,n}$ to add more boundaries giving some of the more intricate structures already displayed. They all come from acting with this operator. For example for ${\widetilde W}_{1,1}$ in equation~(\ref{eq:W11-ell}) was simply made by acting on $F_1{=}{-}\frac{1}{24}\log(u_0^\prime(x))$ once with $\delta^{(u_0)}_{\ell_1}$ and ${\widetilde W}_{1,2}$ in equation~(\ref{eq:W12-ell}) by acting again with $\delta^{(u_0)}_{\ell_2}$.

This completes the whole story of how to write these general formulae at any number of boundaries and genus.

\section{Hard-edge Models and ${\cal N}{=}1$ Supersymmetric applications}
\label{sec:N=1-super-cases}
There is a large class of important applications that deserve some special attention, because their perturbation theory is developed about  $u_0(x){=}0$. This appears to be a challenge for the formulae written so far. However, some careful attention to the matter will reveal  ways to incorporate these cases shortly.    Here is the background as to why these cases arise. 

As mentioned earlier,  there are models for which perturbation theory is developed around the $x{\to}\,{+}\infty$ regime of string equation~(\ref{eq:big-string-equation}), and these are examples of such models. Typically this includes various supersymmetric models, for which the random Hamiltonians under consideration are positive. Indeed, as reviewed in Section~\ref{sec:toolbox}, that string equation can be considered as coming from double-scaling certain Wishart-type models (and multicritical generalizations thereof), which naturally have a lowest energy zero, with a degeneracy~$\Gamma$. Recent work~\cite{Johnson:2023ofr} has shown how to use the solutions of this string equation to study supersymmetric gravity models with extended supersymmetry, with~$\Gamma$ scaling with $N$ such that ${\widetilde\Gamma}{=}\hbar\Gamma$ is finite, yielding a system with ${\widetilde\Gamma}{\sim}{\rm e}^{S_0}$ BPS states, appropriate for black holes with extremal entropy $S_0$. In such a case, the string equation has a non-zero leading  solution $u_0(x)={\widetilde\Gamma}^2/x^2 +\cdots$, a solution of the classical equation:
\begin{equation}
\label{eq:classical-big}
    u_0\left(\sum_{k=1}^\infty t_ku_0^k+x\right)^2={\widetilde\Gamma}^2\ .
\end{equation}
Then, as shown in ref.~\cite{Ahmed:2025lxe}, the structure of perturbation theory is the same as that of the $x{\to}{-}\infty$ regime, and so all equations in this paper written for non-zero $u_0(x)$ apply to such cases.\footnote{The very simplest such model is the  recent case  studied in ref.~\cite{Johnson:2026plw} as a universal topological model
of features of fortuitous BPS chaos. It has {\it all} $t_k=0$ yielding the simple result $u_0(x)={\tilde\Gamma}^2/x^2$. Its spectrum and pattern of $W_{g,n}$ show that it   interpolates between the Bessel model and Airy models in an interesting way.}


However, when $\Gamma$ does not scale with $N$, then the classical limit sends the right hand side of (\ref{eq:classical-big}) to zero, and the solution separates into two branches: ${\cal R}_0[u]{=}0$ applies to the $x{<}0$ regime, giving the familiar type of perturbation theory already discussed, but for $x{>}0$ the leading solution is $u_0(x){=}0$.  For such systems, $\mu$ is  positive, and we are in conventions where $\mu{=}1$. From equation~(\ref{eq:integral-representation}) this yields the leading $\rho_0(E){=}\frac{1}{2\pi\hbar}\frac{1}{\sqrt{E}}+\cdots$ hard wall behaviour typically associated to the Bessel model and various ${\cal N}{=}1$ systems. See refs.~\cite{Johnson:2020heh,Johnson:2020exp} for the first applications of such solutions to the ${\cal N}{=}1$ JT supergravity constructions of ref.~\cite{Stanford:2019vob}.

The consequences of the exact vanishing of $u_0(x)$ for the formalism of the previous sections must be interpreted carefully, given that all our formulae are written in terms of $u_0(x)$.
Indeed, since $u_0(x)$ and all its derivatives actually vanish at any $\mu>0$, any expressions with these quantities purely on the top line will vanish. This includes all tree-level correlation functions beyond $n{=}2$. But for higher genus we should be more careful, since {\it ratios} of $u_0(x)$ derivatives are meaningful finite quantities. 

 The way to see this is to treat the ${\widetilde \Gamma}{=}0$ cases as limits of the situation with ${\widetilde \Gamma}{\neq}0$. It is straightforward to use~(\ref{eq:classical-big}) to write equations for the derivatives of $u_0(x)$ and then evaluate ratios, seeing that they are non-zero even when $u_0(x)$ goes to zero. For example:\begin{eqnarray}
    \label{eq:derivatives}
u_0^{\prime}&=&-\frac{2u_0}{\left[\sum_{k=1}^\infty(2k+1)t_ku_0^k+x\right]}\ , \nonumber\\
u_0^{\prime\prime}&=&\frac{6u_0}{\left[\sum_{k=1}^\infty(2k+1)t_ku_0^k+x\right]^2}\nonumber\\
&&\hskip1cm-4u_0\frac{\sum_{k=1}^\infty(2k+1)kt_ku_0^k}{\left[\sum_{k=1}^\infty(2k+1)t_ku_0^k+x\right]^3}
\ ,
\end{eqnarray}
and so taking the ratio and setting $u_0(x){=}0$ gives $u^{\prime\prime}_0/u^{\prime}_0{=}{-}3/x$. Similar manipulations give some more ratios that  will be useful in a moment, so they are listed here:
\begin{eqnarray}
    \label{eq:derivative-ratios}
    \frac{u^{\prime\prime}_0}{u^{\prime}_0}{=}-\frac{3}{x}\ , \quad
    \frac{u^{(3)}_0}{u^{\prime}_0}=\frac{12}{x^2}\ , \quad
    \frac{u^{(4)}_0}{u^{\prime}_0}=-\frac{60}{x^3}\ . 
\end{eqnarray}
Putting the first ratio into the genus one single-boundary formula~(\ref{eq:W11-z}) gives:
 \begin{equation}
    W_{1,1}(z_1)=-\frac{1}{8z_1^2}\ ,
\end{equation} (after setting $x{=}\mu{=}1$). Looking at the general expression for $W_{1,2}$, equation~(\ref{eq:W12}), most terms vanish since they have just pure powers of $u_0^\prime$. That same combination of ratios as before turns up in the last terms to give:
\begin{equation}
    W_{1,2}(z_1,z_2)=\frac{1}{8z_1^2z_2^2}\ ,
\end{equation}
and finally, looking at general expression~(\ref{eq:W13}), a vast number of terms vanish, except for three which involve surviving derivative ratios again, yielding the combination: $(8\times(-60) + 24\times 36 - 16\times27)/(192 z_1^2z_2^2z_3^2x^3)$, and setting $x{=}\mu{=}1$ gives:
\begin{equation}
    W_{1,3}(z_1,z_2,z_3)=-\frac{1}{4z_1^2z_2^2z_3^2}\ .
\end{equation}
These results are part of a known pattern, proven by Norbury~\cite{Norbury:2020vyi} for these super Weil-Petersson volumes, best written in Laplace transform form:
\begin{equation}
    \label{eq:W1n-norbury}
    V_{1,n}{=}(-1)^n\frac{(n-1)!}{8}\ ,
\end{equation} in a convention where there is an extra factor of $(-)^n$ in front of all supervolumes. Below, we will prove this formula easily with the methods of this paper, as well as a generalization of it.

Other results can be obtained in this way from various conspiracies involving ratios of derivatives of $u_0(x)$, but it is clear that this is not the most efficient language in which to cast the correlators. The simpler form of the $u_{2g}(x)$ offer a more direct route.

The point is  that perturbation theory should be built around developing $u(x){=}0+\hbar^2u_2(x)+\hbar^4u_4(x)+\cdots$ and write new expressions involving the $u_{2g}(x)$. Actually: 
\begin{equation}
\label{eq:u2-x-plus}
    u_2(x)=\frac{\Gamma^2-\frac14}{x^2}\ ,
\end{equation} is in fact universal (no dependence on the~$t_k$ specifying a model), and the string equation~(\ref{eq:big-string-equation}) readily gives the next order in terms of $u_2$ and its derivatives:
\begin{eqnarray}
u_4(x)=\frac{t_1}{x^2}\left[\frac12 xu_2^{\prime\prime}-\frac12 u_2^\prime -2x u_2^2\right]\ ,
\end{eqnarray}
yielding:
\begin{eqnarray}
\label{eq:u4-x-plus}
u_4(x)=-2t_1\frac{\left(\Gamma^2-\frac14\right)\left(\Gamma^2-\frac94\right)}{x^5}\ ,
\end{eqnarray} which is the first place the~$t_k$ dependence occurs, with successive~$t_k$ appearing as the $u_{2g}(x)$ increase in complexity. 

\begin{widetext}
\noindent For later use, here is the next order:
\begin{eqnarray}
u_6(x)
&=&
t_2 \Bigg\{
\frac{1}{x}
\left(
-2u_2^3
+u_2'^2
+\frac{5}{3}u_2 u_2''
-\frac{1}{6}u_2^{(4)}
\right)
+\frac{1}{x^2}
\left(
- u_2 u_2'
+\frac{1}{6}u_2^{(3)}
\right)
\Bigg\}
\\
&+&
t_1^{\,2} \Bigg\{
\frac{1}{x^2}
\left(
7u_2^3
-\frac{9}{4}u_2'^2
-\frac{7}{2}u_2 u_2''
+\frac{1}{4}u_2^{(4)}
\right)
+
\frac{1}{x^3}
\left(
8u_2 u_2'
- u_2^{(3)}
\right)
+
\frac{1}{x^4}
\left(
-3u_2^2
+2u_2''
\right)
+
\frac{1}{x^5}
\left(
-2u_2'
\right)
\Bigg\}\ ,
\nonumber
\end{eqnarray}
which gives:
\begin{eqnarray}
\label{eq:u6-x-plus}
&&u_6(x)=\hbar^6\left(\Gamma^2-\frac14\right)\left(\Gamma^2-\frac94\right)
\left[
\frac{7t_1^2}{x^8}\left(\Gamma^2-\frac{21}{4}\right)
-\frac{2t_2}{x^7}\left(\Gamma^2-\frac{25}{4}\right)
\right]\ .
\end{eqnarray}
Along similar lines, the next order gives:
\begin{eqnarray}
\label{eq:u8-x-plus}
&&u_8(x)
=
\left(\Gamma^2-\frac14\right)
\left(\Gamma^2-\frac94\right)
\times
\\
&&\qquad\qquad
\Bigg[
-\frac{2\,t_3}{x^9}
\left(\Gamma^2-\frac{25}{4}\right)
\left(\Gamma^2-\frac{49}{4}\right)
\nonumber
+\frac{18\,t_1 t_2}{x^{10}}
\left(\Gamma^2-\frac{25}{4}\right)
\left(\Gamma^2-\frac{115}{12}\right)
-\frac{30\,t_1^{3}}{x^{11}}
\left(\Gamma^2-\frac{29}{4}\right)
\left(\Gamma^2-\frac{83}{12}\right)
\Bigg]\ ,
\end{eqnarray}
which will be useful later.
Note that for ${\cal N}{=}1$ JT supergravity  the $t_k$ are given by~\cite{Johnson:2020heh} $t_k {=} \pi^{2k}/(k!)^2$, so  $t_1{=}\pi^2$,  $t_2{=}\pi^4/4$ and $t_3{=}\pi^6/36$.

\end{widetext}
Indeed we see that the derivative ratios~(\ref{eq:derivative-ratios}) yield the right $\Gamma{=}0$ result for $u_2$:
\begin{equation}
    \label{eq:u2-confirm}
     u_2(x)=\frac{u_0''^2-u_0'u_0'''}{12u_0'^2}=\frac{1}{12x^2}\times(9-12)=-\frac{1}{4x^2}\ ,
\end{equation} and a bit more work  with such ratios as they occur in~(\ref{eq:eliminate-u2-u4}) yields correctly that (at $\Gamma{=}0$) $u_4(x){=}{-}9t_1/8x^4$, and so on. A more careful approach to $u_0(x){=}0$ can presumably also get the  $\Gamma\neq0$ results, 
but overall, it is much clearer to work in terms of the~$u_{2g}(x)$ directly rather than appealing to ratios of $u_0(x)$ derivatives.
In fact, a remarkable simplification occurs, because the $u_{2g}(x)$ are merely sums of terms involving inverse powers of $x^2$, and since $u_0(x){=}0$, the $\hbar$~expansion of the Gel'fand-Dikii equation simply gives ${\widehat R}(x,E)$ back in terms of powers of $(-E)^{-\frac12}$ again. The Gel'fand-Dikii polynomials are just polynomials in~$1/x^2$. Everything can be $x$-integrated exactly to give simple expressions for the one-point functions.

To proceed with adding more loops, we just need to use the  reasoning used earlier. When everything was written in terms of $u_0(x)$, all we had to do was use the action of the boundary operator $\delta^{(u_0)}_{E_i}$ on it, which was particularly simple. If we instead leave things in terms of the $u_{2g}(x)$, we can use the action on them given in equation~(\ref{eq:delta-u2g}), but let's write the first form:
\begin{equation}
    \label{eq:delta-u2g-2}
  \delta_{E_i}(u_{2g}(x))=
2\frac{d}{dx}{\widehat R}_g(x,E_i)\ ,
\end{equation}
and  in particular we have:
\begin{eqnarray}
    &&\hskip-1cm
    \delta^{(u_2)}_{E_i}=2\frac{d}{dx}\left[\frac{u_2(x)}{4(-E_i)^\frac32}\right]\ ,\label{eq:delta-u2}\\
    &&\hskip-1cm
    \delta^{(u_4)}_{E_i}={2}\frac{d}{dx}\left[\frac{u_4(x)}{4(-E_i)^\frac32}-\frac{(3u_2(x)^2-u_2^{\prime\prime}(x))}{16(-E_i)^\frac52}\right]\ ,
    \label{eq:delta-u4}
\end{eqnarray}

\begin{widetext}

\noindent with:
\begin{equation}
\label{eq:delta-u6}
\delta^{(u_6)}_{E_i}={2}\frac{d}{dx}\left[\frac{u_6(x)}{4(-E_i)^{\frac32}}
+
\frac{u_4''(x)-6u_2(x)u_4(x)}{16(-E_i)^{\frac52}}
+
\frac{
10u_2(x)^3
-10u_2(x)u_2''(x)
-5\big(u_2'(x)\big)^2
+u_2^{(4)}(x)
}{64(-E_i)^{\frac72}}\right]\ ,
\end{equation}   
and the expression for $\delta^{(u_8)}_{E_i}$ will be given later in~(\ref{eq:delta-u8}), closer to where it will be used.
    
\end{widetext}

Given that $u_0(x)$ and all its derivatives can be set to zero in the expansion of ${\widehat R}_g(x,E_i)$ in equation~(\ref{eq:useful-expansion}), this is another way to see that the volumes for ${\cal N}{=}1$ JT supergravity are quite compact (in comparison to ordinary Weil-Petersson volumes). Let's see how this works, keeping $\Gamma$ general (as is easy to do here) starting with genus~one. (The   meaning of $\Gamma$ will be discussed  just below equation~(\ref{eq:W3n-general}) and in the discussion (Section~\ref{sec:discussion})). We have:
\begin{equation}
    {\widetilde W}_{1,1}(x,E_1)=\frac{1}{4(-E_1)^\frac32}\int^x\!\! u_2(x^\prime) dx^\prime\ ,
\end{equation}
with $u_2(x)$ given in~(\ref{eq:u2-x-plus}). This gives, after identifying $z_1^2=-E_1$,
\begin{equation}
     W_{1,1}(z_1)=\frac12\left(\Gamma^2-\frac14\right)\frac{1}{z_1^2}\ .
\end{equation}
Moving swiftly on, we act with $-\delta^{(u_2)}_{E_2}$ given in~(\ref{eq:delta-u2g-2}) to get:
\begin{equation}
    {\widetilde W}_{1,2}(x,E_1,E_2)=-\frac{2}{4(-E_1)^\frac32\cdot 4(-E_2)^\frac32} u_2(x)\ ,
\end{equation}
which gives:
\begin{equation}
     W_{1,2}(z_1,z_2)=-\frac12\left(\Gamma^2-\frac14\right)\frac{1}{z_1^2z_2^2}\ .
\end{equation}
From here on the pattern is clear, with every further action of $\delta^{(u_2)}_{E_i}$ bringing in a $\frac{1}{2(-E_i)^\frac32}=\frac{1}{2z_i^3}$, but the Jacobian $(-2z_i)$ will convert this into an extra  $\frac{1}{z_i^2}$ factor. Meanwhile, each time an extra derivative is performed on $u_2(x)$ before setting $x{=}1$. But the extra derivative simply multiplies by the next integer, and so we have completed a simple proof of the $\Gamma{=}0$ closed-form result~\cite{Norbury:2020vyi},  for all~$n$, now extended  to general~$\Gamma$:
\begin{eqnarray}
 &&W_{1,n}=(-1)^{n-1}\frac{(n-1)!}{2}\left(\Gamma^2-\frac14\right)\prod_{i=1}^n
 \frac{1}{z_i^2}\ ,\nonumber\\
 \end{eqnarray}
 producing the first of our general volume formulae:
 \begin{eqnarray}
    &&V_{1,n}=(-1)^{n-1}\frac{(n-1)!}{2}\left(\Gamma^2-\frac14\right)\ .
\label{eq:W1n-special}
\end{eqnarray}
Notice that the cases $\Gamma{=}\pm\frac12$ readily yields the known exact vanishing at genus 1 for those models. This will persist to all genus since  $\Gamma^2{-}\frac14$ will persist as a factor.
\\
\\
Next we can look at genus~2, where:
\begin{eqnarray}
\label{eq:W21-special}
    {\widetilde W}_{2,1}(x,E_1)&=&\int^x {\widehat R}_2(x^\prime,E_1)dx^\prime\\&=&\int^x\!\! \left[\frac{u_4(x^\prime)}{4(-E_1)^\frac32}-\frac{(3u_2(x^\prime)^2-u_2^{\prime\prime}(x^\prime))}{16(-E_1)^\frac52}\right] dx^\prime\ ,\nonumber
\end{eqnarray}
with $u_2(x)$ given in~(\ref{eq:u2-x-plus}), and $u_4(x)$  in~(\ref{eq:u4-x-plus}). This integrates readily to give:
\begin{eqnarray}
    &&{\widetilde W}_{2,1}(E_1)=\left(\Gamma^2-\frac14\right)\left(\Gamma^2-\frac94\right)\times\\
    &&\hskip3cm\left[\frac{t_1}{8(-E_1)^\frac32}+\frac{1}{16(-E_1)^\frac52}\right]\ ,\nonumber
\end{eqnarray}
which yields:
\begin{equation}
    \nonumber\\
   V_{2,1}=-\left(\Gamma^2-\frac14\right)\left(\Gamma^2-\frac94\right)\left[\frac{t_1}{4}+\frac{b_1^2}{48}\right] \ ,
\end{equation}
and for ordinary ${\cal N}{=}1$ supersymetric JT gravity ($\Gamma{=}0,t_1{=}\pi^2$)
this gives the known result (see {\it e.g.,} Appendices in ref.~\cite{Fuji:2023wcx}):
\begin{equation}
\label{eq:N=1V21}
    V_{2,1}=-\left(\frac{9\pi^2}{64}+\frac{3b_1^2}{256}\right)\ .
\end{equation}

Now  for the first non-trivial test of our modified procedure for adding loops. Since both $u_2$ and $u_4$ enter into (\ref{eq:W21-special}) we must act with  both (\ref{eq:delta-u2}) and (\ref{eq:delta-u4}) on it (with the overall extra $(-1)$), giving the pleasing result:

\begin{widetext}
\begin{eqnarray}
  \hskip-0.4cm {\widetilde W}_{2,2}(x,E_1,E_2)=-\Biggl[
\frac{u_4(x)}{8(-E_1)^\frac32(-E_2)^\frac32}-\frac{(3u^2_2(x)-u_2^{\prime\prime}(x))}{32}\left(
\frac{1}{(-E_1)^\frac52(-E_2)^\frac32}+\frac{1}{(-E_2)^\frac52(-E_1)^\frac32}\right)\Biggr] \ ,
\label{eq:W22-special}
\end{eqnarray}
Note here that we could have written everything in terms of $u_2(x)$ and its derivatives, thereby matching what was done for $u_0(x)$ in the rest of the paper, but since things are so simple here, it is just as efficient to present things in terms of the $u_{2g}(x)$. (As we go to higher genus, more of them will become involved in a quite natural way: At genus~$g$, $u_{2g}(x)$ and all lower $u_{2g'}(x)$ ($g'<g$) will be in play.)
In any case,  at $x=\mu=1$ the result is:
\begin{eqnarray}
  \hskip-0.4cm {\widetilde W}_{2,2}(E_1,E_2)=\left(\Gamma^2-\frac14\right)\left(\Gamma^2-\frac94\right)\left(\frac{t_1}{4(-E_1)^\frac32(-E_2)^\frac32}+
\frac{3}{32(-E_1)^\frac52(-E_2)^\frac32}+\frac{3}{32(-E_2)^\frac52(-E_1)^\frac32}\right) \ ,
\end{eqnarray}
which after inserting the Jacobian factor $(-2z_1)(-2z_2)$ is the Laplace transform of the known result at ($\Gamma=0,t_1=\pi^2)$:
    \begin{equation}\label{eq:N=1V22}
        V_{2,2}=\left(\frac{9 \pi^{2}}{16}+\frac{9 b_{1}^{2}}{256}+\frac{9 b_{2}^{2}}{256}\right)\ .
    \end{equation}
This result and the case~(\ref{eq:N=1V21}) are  special cases of a more general known formula, again by Norbury~\cite{Norbury:2020vyi}:
\begin{equation}
\label{eq:N=1V2n}
    V_{2,n}=(-1)^n
\frac{3 (n+1)!}{2 \cdot 4^4}
\left[
(2\pi)^2 (n+2)
+ \sum_{i=1}^{n} b_i^2
\right]\ .
\end{equation}
Just as with genus 1, it is straightforward to see how the loop operator action generates this. First, it is best to Laplace transform~(\ref{eq:N=1V2n}) to:
\begin{equation} 
\label{eq:N=1W2n}
W_{2,n}(\{z_n\})
=(-1)^n
\frac{3}{2}\,\frac{(n+1)!}{4^4}
\left(\prod_{i=1}^n \frac{1}{z_i^2}\right)
\left((2\pi)^2 (n+2)
+ 6 \sum_{i=1}^n \frac{1}{z_i^2}\right)\ .
\end{equation}

The step that got from~(\ref{eq:W21-special}) to~(\ref{eq:W22-special}) by acting with $\delta_{E_2}^{(u_4)}$ and $\delta_{E_2}^{(u_2)}$ has a distinctive pattern to it. It guarantees that the resulting dependence on the $(-E_i)^{-\frac12}$ is a symmetric polynomial in them, and repeated application to add more boundaries produces as coefficients the {\it same} combinations of functions $u_4$ and $u_2$, but differentiated. After some practice, the pattern is, writing $z_i=(-E_i)^\frac12$ for short:
\begin{eqnarray}
  \hskip-0.4cm {\widetilde W}_{2,n}(x,\{z_i\})=-\frac{2^n}{2\cdot 4^n}\prod_{i=1}^n \frac{1}{z_i^3}
  \left\{
\left[u_4(x)\right]^{(n-2)}-\frac14\sum_{i=1}^n\frac{1}{z_i^2} \left[3u^2_2(x)-u_2^{\prime\prime}(x)\right]^{(n-2)}\right\} \ ,
\label{eq:W2n-special}
\end{eqnarray}
where the superscript $(n-2)$ means differentiate $(n-2)$ times.
That the answer just involved $n-2$ derivatives of our simple functions means that the $n$-dependence is easy to extract, since all $u_{2g}(x)$ are just built from sums of powers of $1/x$. For this case see ~(\ref{eq:u2-x-plus}) and (\ref{eq:u4-x-plus}). Differentiating and setting $x=\mu=1$ gives a nice general result (throwing in Jacobian $\prod_i^n(-2z_i)$):
\begin{eqnarray}
  \hskip-0.4cm W_{2,n}(\{z_i\})&=&(-)^{n-1}\frac12\left(\Gamma^2-\frac14\right)\left(\Gamma^2-\frac94\right)\prod_{i=1}^n \frac{1}{z_i^2}
  \left\{
-2t_1\frac{(n+2)!}{24}-\frac34\frac14\sum_{i=1}^n\frac{1}{z_i^2} \frac{(n+1)!}{3!}\right\}\nonumber\\
&=&(-)^n\frac{(n+1)!}{6\cdot4^2}\left(\Gamma^2-\frac14\right)\left(\Gamma^2-\frac94\right)\prod_{i=1}^n \frac{1}{z_i^2}
  \left\{
4t_1(n+2)+6\sum_{i=1}^n\frac{1}{z_i^2} \right\}
\ ,
\label{eq:W2n-special}
\end{eqnarray}
which reduces to that of Norbury when $\Gamma=0$ and $t_1=\pi^2$. (Notice that there is now vanishing at $\Gamma{=}\pm\frac32$, in addition to the $\Gamma{=}\pm\frac12$ cases. As also noted in ref.~\cite{Johnson:2020heh}, it would be interesting to understand these new models, and other half-integer cases where special features occur, from a gravity perspective.)

Emboldened by this success, it is natural to turn to genus three, for which there is the last of Norbury's closed-form formulae~\cite{Norbury:2020vyi}:
\begin{equation}
    V_{3,n}=
\frac{(n+3)!}{5 \cdot 4^8}\,(-1)^n
\left[
(2\pi)^4 (n+4)(42n+185)
+ 84(2\pi)^2 (n+4)\sum_{i=1}^{n} b_i^2
+ 25\sum_{i=1}^{n} b_i^4
+ 84\sum_{1\leq i< j \leq n} b_i^2 b_j^2
\right]\ .
\end{equation}
    Once again, we will see the underlying pattern in this formula can be attributed to the action of our loop operator. The better target is the Laplace transform:
\begin{equation}
W_{3,n}
=
\frac{(n+3)!}{5\cdot 4^8}\,(-1)^n
\left(\prod_{k=1}^{n} \frac{1}{z_k^2}\right)
\Bigg[
(2\pi)^4 (n+4)(42n+185)
+ 504(2\pi)^2 (n+4)\sum_{i=1}^{n} \frac{1}{z_i^2}
+ 3000\sum_{i=1}^{n} \frac{1}{z_i^4}
+ 3024\sum_{1\leq i< j \leq n} \frac{1}{z_i^2 z_j^2}
\label{eq:N=1W3n}
\Bigg].
\end{equation}
As before, we start with the basic result, the integrated Gel'fand-Dikii resolvent at order $g{=}3$, now also involving  $u_6$:
\begin{eqnarray}
   &&{\widetilde W}_{3,1}(x,E_1)=
   \int^x\Biggl( \frac{u_{6} (x)}{4 \left(-E_1 \right)^{\frac{3}{2}}}+\frac{1}{16\left(-E_1 \right)^{\frac{5}{2}}}
\left({u^{\prime\prime}_{4}(x)}-{6 u_{2}(x) u_{4}(x)}\right)
    \nonumber\\
&&\hskip5cm    +\frac{1}{64\left(-E_1 \right)^{\frac{7}{2}}}\left({10 u_{2}(x)^{3}}-{10 u_{2}(x) u^{(2)}_{2}\!(x)}-{10 \left(u^\prime_{2}(x)\right)^{2}}+{u^{(4)}_{2}\!(x)}\right)\Biggr) dx\ ,\nonumber
\end{eqnarray} which readily gives the known $W_{3,1}$ upon the usual substitutions~\cite{Johnson:2024bue}. To add boundaries we act with the same $\delta_{E_i}^{(u_4,u_2)}$ as before, but additionally $\delta^{(u_6)}_{E_i}$ in (\ref{eq:delta-u6}). Acting once under the integral sign gives a total derivative as before, which readily integrates to:
\begin{equation}
    {\widetilde W}_{3,2}(x,z_1,z_2) = 
-\prod_{i=1}^{2}\frac{1}{z_i^3}\,
\left[
A_{32}(x)
+ B_{32}(x)\,\sum_{i=1}^{2}\frac{1}{z_i^2}
+ C_{32}(x)\,\sum_{i=1}^{2}\frac{1}{z_i^4}
+ D_{32}(x)\!\!\sum_{1\le i<j\le 2}\frac{1}{z_i^2 z_j^2}
\right]\ ,
\end{equation}
where (noting the overall minus sign):
\begin{align}
\label{eq:ABCDW32}
A_{32}\;&=\;\frac{1}{8}\,u_{6}(x),\\[4pt]
B_{32}\;&=\;- \frac{3}{8}\,u_{2}(x)\,u_{4}(x) + \frac{1}{16}\,u_{4}''(x),\\[4pt]
C_{32}\;&=\;\frac{5}{32}\,u_{2}(x)^3 -  \frac{5}{32}\,u_{2}''(x)\,u_{2}(x) -  \frac{5}{64}\,u_{2}'(x)^2 + \frac{1}{64}\,u_{2}''''(x),\\[4pt]
D_{32}\;&=\;\frac{3}{32}\,u_{2}(x)^3 -  \frac{3}{32}\,u_{2}''(x)\,u_{2}(x) -  \frac{3}{128}\,u_{2}'(x)^2 + \frac{1}{128}\,u_{2}''''(x) \ .
\end{align}
This case is somewhat  unwieldier than the previous case, but  the same kinds of patterns emerge:  symmetric polynomials in the $1/z_i$ result after every action of the boundary insertion, and remarkably, after the $n$th
 action, $(n-2)$ derivatives of the {\it same} combinations of functions seen in $\widetilde{W}_{3,2}$ appear as the coefficients!\footnote{\label{fn:AI-comment}OpenAI's  {\tt ChatGPT5.2} was a very helpful tool for organizing and searching the many-pages-long expressions generated in {\tt Maple} using the loop operator, helping confirm the pattern for cases $n{=}3,4,5$ and $6$.}
The general pattern that emerges is:
\begin{equation}
{\widetilde W}_{3,n}(x,\{z_n\}) = (-1)^n
\prod_{i=1}^{n}\frac{1}{z_i^3}\,
\left[
A_{3n}(x)
+ B_{3n}(x)\sum_{i=1}^{n}\frac{1}{z_i^2}
+ C_{3n}(x)\sum_{i=1}^{n}\frac{1}{z_i^4}
+ D_{3n}(x)\!\!\!\sum_{1\le i<j\le n}\frac{1}{z_i^2 z_j^2}
\right]\ ,
\end{equation}
and, remarkably:
\begin{equation}
\label{eq:ABCDW3n}
    A_{3n}=\left(\frac{1}{2}\frac{\partial}{\partial x}\right)^{(n-2)}\!\!\! A_{32}\ ,\quad B_{3n}=\left(\frac{1}{2}\frac{\partial}{\partial x}\right)^{(n-2)}\!\!\! B_{32}\ ,\quad C_{3n}=\left(\frac{1}{2}\frac{\partial}{\partial x}\right)^{(n-2)}\!\!\! C_{32}\ ,\quad D_{3n}=\left(\frac{1}{2}\frac{\partial}{\partial x}\right)^{(n-2)}\!\!\! D_{32}\ .
\end{equation}
Once again since the coefficient functions in (\ref{eq:ABCDW32}) and (\ref{eq:ABCDW3n}) are simply built from functions ($u_2,u_4,u_6,u_8$) that are in turn polynomials in $1/x$, it is straightforward to write the general $n$ dependence that results! After putting in the same Jacobian as before, and some tidying, the final general result is:
\begin{align}
\label{eq:W3n-general}
W_{3,n}(\{z_i\})=&(-1)^{n+1}\frac{ (n+3)!}{5760}\,
\Bigl(\Gamma^2-\frac94\Bigr)\Bigl(\Gamma^2-\frac14\Bigr)\prod_{i=1}^{n}\frac{1}{z_i^2}
\Bigg[
4(n+4)
\Bigl(
(n+5)\Bigl(\Gamma^2-\frac{21}{4}\Bigr)t_1^2
-2\Bigl(\Gamma^2-\frac{25}{4}\Bigr)t_2
\Bigr)
\nonumber\\[4pt]
&\qquad\qquad
+12(n+4)\Bigl(\Gamma^2-\frac{21}{4}\Bigr)t_1\,\sum_{i=1}^{n}\frac{1}{z_i^2}
+15\Bigl(\Gamma^2-\frac{25}{4}\Bigr)\sum_{i=1}^{n}\frac{1}{z_i^4}
+18\Bigl(\Gamma^2-\frac{21}{4}\Bigr)\sum_{1\le i<j\le n}\frac{1}{z_i^2 z_j^2}
\Bigg]\ ,
\end{align}
which not only reduces to Norbury's formula (\ref{eq:N=1W3n}) (at $\Gamma{=}0$, $t_1{=}\pi^2$ and $t_2{=}\pi^4/4$), but captures more general cases too. 

For all the formulae derived in this section, in addition to allowing arbitrary $t_k$, $\Gamma$ was general too. For this case and the next one to come, it is also noticeable that there are other special values of $\Gamma$ that yield interesting physics (besides the well-known $\Gamma{=}0$ and $\Gamma{=}{\pm}\frac12$ cases of ref.~\cite{Stanford:2019vob}). Recall  that $\Gamma$ parameterizes a model in the  $(2\Gamma+1,2)$ Altland-Zirnbauer classification~\cite{Altland:1997zz}. These  should be considered as all distinct models, for integer and half-integer $\Gamma$. 

There is another (quite exciting from some perspectives) interpretation of $\Gamma$ that is worth noting, that allows these formulae to be read a different way, giving expressions for supersymmetric Weil-Petersson volumes with additional {\it Ramond} boundaries\cite{Witten:2012ga} (the presentation here has implicitly been about just  {\it Neveu-Schwarz} boundaries). This is explored in the Discussion (Section~\ref{sec:discussion}), and in a forthcoming paper~\cite{Johnson:2026jls}.

The relative ease with which we were able to build useful closed-form formulae for genus three and below suggests that such formulae are possible for any genus, and moreover that it is straightforward to derive entirely new ones.
As an illustration, it is worth doing genus four, for which  the function $u_8(x)$, and hence $t_3$ (see~(\ref{eq:u8-x-plus})), enters the fray. What will be needed additionally is $\delta^{(u_8)}_{E_i}$, the loop addition operator's action on $u_8$, which is:
\begin{eqnarray}
\label{eq:delta-u8}
&&\hskip-0.5cm\delta^{(u_8)}_{E_i}={2}\frac{d}{dx}\Biggl[
\frac{u_8(x)}{4\,(-E_i)^{3/2}}
+\frac{u_6''(x)-6\,u_2(x)u_6(x)-3\,u_4(x)^2}{16\,(-E_i)^{5/2}}
\\\nonumber
&&\hskip2cm
+\frac{u_4^{(4)}(x)-10\,u_2(x)u_4''(x)-10\,u_2'(x)u_4'(x)-10\,u_2''(x)u_4(x)+30\,u_2(x)^2u_4(x)}{64\,(-E_i)^{7/2}}
\\\nonumber
&&\quad
+\frac{70\,u_2(x)^2u_2''(x)+70\,u_2(x)\bigl(u_2'(x)\bigr)^2-14\,u_2(x)u_2^{(4)}(x)-28\,u_2'(x)u_2^{(3)}(x)-35\,u_2(x)^4+u_2^{(6)}(x)-21\,\bigl(u_2''(x)\bigr)^2}{256\,(-E_i)^{9/2}}
\Biggr]\ ,\nonumber
\end{eqnarray} 
where recall that the content of the square bracket is simply ${\widehat R}_4(x,E_i)$, the order $\hbar^{8}$ correction to the Gel'fand-Dikii equation solution. Correspondingly, we start with the genus four one-boundary object:
\begin{equation}
    {\widetilde W}(x,E_1)=\int^x {\widehat R}_4(x',E_i) dx'\ , 
\end{equation} and repeatedly act with ${(-)}\delta_{E_i}$ to add the $i$th boundary.
The natural guess is that the structures  seen for the lower genus cases will persist. Confirming that and extracting the repeating structures that get multiply differentiated (the analogues of the genus three objects in (\ref{eq:ABCDW32})
is difficult to do given the length of the expressions that result even after a couple of iterations. This is where the asset mentioned in  footnote~\ref{fn:AI-comment} was no longer just a convenience but an essential tool. The structure that is confirmed is:
\[
W_{4,n}
=
(-1)^{n}\left(\prod_{i=1}^{n}\frac{1}{z_i^{3}}\right)
\Bigg[
A_{4n}(x)
+ B_{4n}(x)\sum_{i=1}^{n}\frac{1}{z_i^{2}}
+ C_{4n}(x)\sum_{i=1}^{n}\frac{1}{z_i^{4}}
+ D_{4n}(x)\sum_{1\le i<j\le n}\frac{1}{z_i^{2} z_j^{2}}
\]
\[
\hspace{2.7cm}
+ E_{4n}(x)\sum_{i=1}^{n}\frac{1}{z_i^{6}}
+ F_{4n}(x)\!\!\!\sum_{\substack{1\le i,j\le n\\ i\ne j}}\frac{1}{z_i^{4} z_j^{2}}
+ G_{4n}(x)\!\!\!\sum_{1\le i<j<k\le n}\frac{1}{z_i^{2} z_j^{2} z_k^{2}}
\Bigg],
\qquad\ ,
\]
where for $n=2$, the last term is identically zero (since there's no way of making such a term with the condition stated in the sum), but it is important at higher order, as will emerge below. The coefficients at $n{=}2$ are: 
\begin{align}
\nonumber
A_{42}(x) &= \frac{1}{8}u_{8}(x)\ ,\\
\nonumber
B_{42}(x) &= \frac{1}{32} u_{6}''(x)
-\frac{3}{16} u_{2}(x) u_{6}(x)
-\frac{3}{32} u_{4}(x)^{2}\ ,\\
\nonumber
C_{42}(x) &= \frac{1}{128} u_{4}^{(4)}(x)
-\frac{5}{64} u_{2}(x) u_{4}''(x)
-\frac{5}{64} u_{4}(x) u_{2}''(x)
-\frac{5}{64} u_{2}'(x) u_{4}'(x)
+\frac{15}{64} u_{2}(x)^{2} u_{4}(x)\ ,\\
D_{42}(x) &= \frac{1}{128} u_{4}^{(4)}(x)
-\frac{3}{32} u_{2}(x) u_{4}''(x)
-\frac{3}{32} u_{4}(x) u_{2}''(x)
-\frac{3}{64} u_{2}'(x) u_{4}'(x)
+\frac{9}{32} u_{2}(x)^{2} u_{4}(x)\ ,
\end{align}
with:
\begin{align}
\nonumber
E_{42}(x) &= \frac{1}{512} u_{2}^{(6)}(x)
-\frac{7}{256} u_{2}(x) u_{2}^{(4)}(x)
-\frac{7}{128} u_{2}'(x) u_{2}^{(3)}(x)
-\frac{21}{512} \big(u_{2}''(x)\big)^{2}
\\
&\hskip5cm
+\frac{35}{256} u_{2}(x)^{2} u_{2}''(x)
+\frac{35}{256} u_{2}(x) \big(u_{2}'(x)\big)^{2}
-\frac{35}{512} u_{2}(x)^{4}\ ,
\end{align}
and
\begin{align}
\nonumber
F_{42}(x) &= \frac{1}{512} u_{2}^{(6)}(x)
-\frac{1}{32} u_{2}(x) u_{2}^{(4)}(x)
-\frac{3}{64} u_{2}'(x) u_{2}^{(3)}(x)
-\frac{23}{512} \big(u_{2}''(x)\big)^{2}\\
&\hskip5cm 
+\frac{45}{256} u_{2}(x)^{2} u_{2}''(x)
+\frac{15}{128} u_{2}(x) \big(u_{2}'(x)\big)^{2}
-\frac{45}{512} u_{2}(x)^{4}\ .
\end{align}
The remarkable thing is that all higher $A_{4n}, B_{4n},C_{4n},\ldots$  are {\it again} obtained by acting $(n-2)$ times with  $\frac12\frac{d}{dx}$. The exception is that $G_{4n}$ starts at $n{=}3$ and is:
\begin{equation}
\begin{aligned}
G_{43}(x)
={}&-\frac{27}{128} u_{2}(x)^{3} u_{2}'(x)
+\frac{27}{256} u_{2}(x)^{2} u_{2}^{(3)}(x)
+\frac{81}{256} u_{2}(x) u_{2}'(x) u_{2}''(x)
-\frac{9}{512} u_{2}(x) u_{2}^{(5)}(x)
\\
&\quad
+\frac{9}{256} \big(u_{2}'(x)\big)^{3}
-\frac{9}{256} u_{2}'(x) u_{2}^{(4)}(x)
-\frac{9}{128} u_{2}''(x) u_{2}^{(3)}(x)
+\frac{1}{1024} u_{2}^{(7)}(x).
\end{aligned}
\end{equation}
and hence subsequent $G_{4n}$ come from acting on this $n{-}3$ times with 
$\frac12\frac{d}{dx}$. From here on it is plain sailing, since again all $x$ dependence is in polynomials in $1/x$ and so general $n$ expressions for the derivatives are straightforward to write. After a bit of algebra and tidying, the resulting genus 4 closed-form formula  is ({\it i.e.,} $\Gamma{=}0,t_1=\pi^2,t_2{=}\frac{\pi^4}{4},t_3=\frac{\pi^6}{36}$): 
\begin{equation}
\begin{aligned}
W_{4,n}=&\frac{(-1)^{n}(n+5)!}{315\cdot 2^{19}}\left(\prod_{i=1}^{n}\frac{1}{z_i^{2}}\right)
\Bigg[(n+6)\pi^6(19256n^2+254340n+841736)+
(n+6)\pi^4(86652n+554814)\sum_{i=1}^{n}\frac{1}{z_i^{2}}  
\\ 
&\qquad\qquad\qquad
+258750(n+6)\pi^2  \sum_{i=1}^{n}\frac{1}{z_i^{4}}
+259956(n+6)\pi^2 \sum_{1\le i<j\le n}\frac{1}{z_i^{2} z_j^{2}}
+385875 \sum_{i=1}^{n}\frac{1}{z_i^{6}}\\&\qquad\qquad\qquad\qquad
+388125 \sum_{\substack{1\le i,j\le n\\ i\ne j}}\frac{1}{z_i^{4} z_j^{2}}
+389934 \sum_{1\le i<j<k\le n}\frac{1}{z_i^{2} z_j^{2} z_k^{2}}
\Bigg]\ ,
\end{aligned}
\end{equation}
resulting in a closed-from result the ${\cal N}=1$ supersymmetric Weil-Petersson (NS) volumes for genus four:
\[
\begin{aligned}
{}&V_{4,n}
=
\frac{(-1)^n (n+5)!}{315\,2^{19}}
\Bigg[
(n+6)\pi^6(19256n^2+254340n+841736)
+(n+6)\pi^4(14442n+92469)\sum_{i=1}^n b_i^2
\\
&\qquad\quad
+\frac{8625}{4}(n+6)\pi^2\sum_{i=1}^n b_i^4
+7221(n+6)\pi^2
\!\!\!\!\sum_{1\le i<j\le n} 
\!\!\!\!b_i^2 b_j^2
+\frac{8575}{112}\sum_{i=1}^n b_i^6
+\frac{8625}{16}\sum_{\substack{1\le i,j\le n\\ i\ne j}} b_i^4 b_j^2
+\frac{7221}{4}\!\!\!\!\sum_{1\le i<j<k\le n} \!\!\!\!b_i^2 b_j^2 b_k^2
\Bigg]\ ,
\end{aligned}
\]
which seems   to be new to the literature. (It can be checked that it yields known results for $V_{4,n}$ for some low $n$ cases, for example in ref.~\cite{Fuji:2023wcx}.)

The full general $W$-formula (general $\Gamma,\{t_k\}$) is: 
  \begin{align}
  \nonumber
& W_{4,n}={(-1)^{n}(n+5)!} (4\Gamma^2-9)(4\Gamma^2-1)\left(\prod_{i=1}^{n}\frac{1}{z_i^{2}}\right)\times
\\\nonumber
&\hskip1.0cm 
\Bigg[
\frac{n+6}{185794560}\Big(
(n+7)(n+8)(12\Gamma^2-83)(4\Gamma^2-29) t_1^3\\\nonumber
&\hskip2.5cm
-6(n+7)(12\Gamma^2-115)(4\Gamma^2-25) t_1t_2
+18(4\Gamma^2-49)(4\Gamma^2-25) t_3
\Big)
\\\nonumber &\hskip1.5cm+\frac{n+6}{41287680}
\Big(
(n+7)(4\Gamma^2-29)(12\Gamma^2-83) t_1^2
-2(12\Gamma^2-115)(4\Gamma^2-25) t_2
\Big) \sum_{i=1}^{n}\frac{1}{z_i^{2}}
\\
&\nonumber \hskip1.5cm
+\frac{n+6}{16515072} t_1(12\Gamma^2-115)(4\Gamma^2-25)  \sum_{i=1}^{n}\frac{1}{z_i^{4}}
+\frac{n+6}{13762560} t_1(12\Gamma^2-83)(4\Gamma^2-29) \sum_{1\le i<j\le n}\frac{1}{z_i^{2} z_j^{2}}
\\\nonumber
&\hskip1.5cm
+\frac{1}{4718592}(4\Gamma^2-49)(4\Gamma^2-25) \sum_{i=1}^{n}\frac{1}{z_i^{6}}
+\frac{1}{11010048}(12\Gamma^2-115)(4\Gamma^2-25) \sum_{\substack{1\le i,j\le n\\ i\ne j}}\frac{1}{z_i^{4} z_j^{2}}
\\
&\hskip4cm
+\frac{1}{9175040}(12\Gamma^2-83)(4\Gamma^2-29) \sum_{1\le i<j<k\le n}\frac{1}{z_i^{2} z_j^{2} z_k^{2}}
\Bigg].
\label{eq:W4n-general}
\end{align}
The resulting general volume formula is:
\begin{align}
V_{4,n}(b_1,\dots,b_n)
={}&
(-1)^n (n+5)! \,(4\Gamma^2-9)(4\Gamma^2-1)\times\\
{}&\hskip-0.5cm\Bigg[
\frac{n+6}{185794560}\Big(
(n+7)(n+8)(12\Gamma^2-83)(4\Gamma^2-29)\, t_1^3
\nonumber\\
&\hspace{2.0cm}
-6(n+7)(12\Gamma^2-115)(4\Gamma^2-25)\, t_1 t_2
+18(4\Gamma^2-49)(4\Gamma^2-25)\, t_3
\Big)
\nonumber\\[1mm]
&
+\frac{n+6}{247726080}
\Big(
(n+7)(4\Gamma^2-29)(12\Gamma^2-83)\, t_1^2
-2(12\Gamma^2-115)(4\Gamma^2-25)\, t_2
\Big)\sum_{i=1}^{n} b_i^{2}
\nonumber\\[1mm]
&
+\frac{n+6}{1981808640}\,t_1(12\Gamma^2-115)(4\Gamma^2-25)
\sum_{i=1}^{n} b_i^{4}
+\frac{n+6}{495452160}\,t_1(12\Gamma^2-83)(4\Gamma^2-29)
\sum_{1\le i<j\le n} b_i^{2} b_j^{2}
\nonumber\\[1mm]
&
+\frac{1}{23781703680}(4\Gamma^2-49)(4\Gamma^2-25)
\sum_{i=1}^{n} b_i^{6}
+\frac{1}{7927234560}(12\Gamma^2-115)(4\Gamma^2-25)
\sum_{\substack{1\le i,j\le n\\ i\ne j}} b_i^{4} b_j^{2}
\nonumber\\[1mm]
&
+\frac{1}{1981808640}(12\Gamma^2-83)(4\Gamma^2-29)
\sum_{1\le i<j<k\le n} b_i^{2} b_j^{2} b_k^{2}
\Bigg]\ ,
\end{align}
which presumably (in view of remarks in the discussion section (next) contains information about volumes with Ramond punctures as well.)
 It is quite striking how straightforward it is to derive  these closed forms with this method. The pattern for how they emerge is  likely persists to all genus, so it is natural to conjecture that there is a simple closed form for all $V_{g,n}$ ($W_{g,n}$).

One might wonder if this method can give easy access to closed forms for $W_{g,n}$ ($V_{g,n}$) for general $g$ too. However, notice that as $g$ grows, the overall degree of the symmetric polynomials allowed can increase, allowing new kinds of term not seen at lower $g$ to appear. In addition, at a given $g$, for large enough $n$, new kinds of term analogous to the  $G_{4n}$ for $n>3$ can begin to appear. This is all straightforward to accommodate at a given $g$, but likely makes useful closed form formulae across $g$
 much harder to construct.

\newpage

\end{widetext}

\section{Discussion}
\label{sec:discussion}

Once the dust settles, the construction is all rather simple. The main object doing the work is the loop operator $\delta_{E_i}$, and the key point is that when acting on functions of $u_0(x)$, it dramatically simplifies to (either in energy variables or length variables) $\delta^{(u_0)}_{E_i}$, an operator that yields new~$u_0(x)$ (and $E_i$) dependence, making the result ready to act on  again to add another boundary. 
In the case where $u_0(x){=}0$, one instead works with one or more higher $u_{2g}(x)$, using the simple form for $\delta_{E_i}^{(u_{2g})}$ as needed, and the same remarks apply.

In this way, it is straightforward to derive very swiftly a formula for
any $W_{g,n}$! From this perspective, this makes the method quite attractive because: (1) It is model independent: the result for any $W_{g,n}$ does not need detailed information about the content of $u_0(x)$, so the formulae are quite general. (2) Differentiating, which is the primary computational tool here,  is quick and easy.

The simplicity of  $\delta^{(u_0)}_{E_i}$ is at the heart why ref.\cite{Johnson:2024bue}'s ODE approach  was so powerful, and its helpfulness continued here in this paper: The Gel'fand-Dikii equation, when supplemented with the requirement that~$u(x)$ solves the string equation, generated ${\widehat R}(E_1,x)$  neatly as a total derivative  at every $g>1$, which readily became the $W_{g,1}(E_1)$. As we saw, the total derivative structure itself can be traced to the fact that $\delta^{(u_0)}_{E_1}$ acted directly on the free energy to produce $W_{g,1}(E_1)$.
Moreover, the simple action of the $\delta_{E_i}^{(u_{2g})}$ is given by the perturbative expansion of the ODE.


As stated in the introduction, this paper could well have been written in 1991 or thereabouts, since some the key components (the KdV flows, the loop operator in term of them, {\it etc}.) had been known since that time. Of course, it wasn't, and moreover the many connections that motivate developing these results and writing paper now were not to emerge until the decades to follow. 

Keeping our eye on that period for a while longer is interesting nonetheless. A key output of this work
is that there are strikingly direct relations between the $W_{g,n}$ through the action of the loop operator (at a given $g$) or across $g$ {\it via} the genus expansion of ODEs. Such relations are hard to see in approaches more directly tied to the geometry of ${\cal M}_{g,n}$. It is the underlying KdV flows that enable these relations to be easily derived.

But of course this is very much in the spirit of what transpired with Kontsevich-Witten theory back in 1990/1991. Witten conjectured~\cite{Witten:1990hr} that the generating function of  correlators of pointlike operators corresponding to intersection numbers is   a $\tau$-function of KdV, which is related to the free energy used here as  $F{=}\log \tau$. Kontsevich proved the conjecture directly using a special matrix model~\cite{Kontsevich:1992ti}. The framework presented here shows how all these data are packaged into $W_{g,n}(\{z_i\})$ using~$u_0$ and its derivatives, and showed that the resulting formulae work {\it universally} for a wide class of models (not just bosonic topological gravity): minimal strings (which are $\kappa$-class deformations of the topological gravity case,) Weil-Petersson volumes, and a variety of supersymmetric generalizations of all these. 

The latter examples deserve some additional remarks for another reason: From perusing the mathematics literature it becomes  apparent that  it is highly non-trivial that the ${\cal N}{=}1$ super supersymmetric case also involves a KdV $\tau$-function, and a lot of hard work has gone into making that robust (see {\it e.g.,} refs.~\cite{Norbury:2017eih,Norbury:2020vyi}). By contrast, it is straightforward and natural using the approach taken here (which began, in this context, with papers~\cite{Johnson:2020heh} and \cite{Johnson:2020exp}): The  KdV $\tau$-function structure is built in, and one simply changes the function $u(x)$ that is in play, by just changing the governing string equation from~(\ref{eq:little-string-equation}) to~(\ref{eq:big-string-equation}). In the mathematical literature this statement about the string equation solution amounts to~$\tau$ now being a Brezin-Gross-Witten $\tau$-function, which is more commonly associated with phenomena in Unitary matrix models~\cite{Gross:1980he,Brezin:1980rk}. However,  that relationship at the level of string equations had already been  made clear back in 1992 with the explicit maps presented in ref.~\cite{Dalley:1992br}.  Given that subsequent work has also shown how to incorporate extended supersymmetry with this same string equation~\cite{Johnson:2023ofr,Johnson:2024tgg,Johnson:2025oty,Ahmed:2025lxe}, this immediately predicts that the ${\cal N}{=}2$ and ${\cal N}{=}4$ cases, including all the interesting properties of their 
Weil-Petersson volumes, are again controlled by a KdV $\tau$-function, and it is natural to conjecture that the precise underlying mathematical framework will eventually bear this out. 

Further along those lines is the matter of writing ${\cal N}{=}1$ volumes in the presence of  {\it Ramond } boundaries and punctures~\cite{Witten:2012ga}. This is again a delicate matter when tackled directly in terms of the geometry of supermanifolds. The results for various special cases that have been worked out in ref.~\cite{norbury2024superweilpeterssonmeasuresmoduli,Alexandrov:2024kuj} are actually readily obtainable from the general $V_{g,n}$ (or~$W_{g,n}$) formulae ({\it i.e.,} containing $\Gamma$) developed here in Section~\ref{sec:N=1-super-cases}. The point is that $\hbar\Gamma$ counts a Ramond  insertion.  The fact that only powers of $\Gamma^2$ emerge in the $u(x)$ expansion (and hence in the volumes) corresponds directly to the fact that only even numbers of punctures give non-zero results.\footnote{These features also have a  direct  understanding in the context of how the  string equation emerges in the type~0A minimal string work of ref.~\cite{Klebanov:2003wg}.} Pulling in  a  factor of $\hbar$ into every $\Gamma$ that appears changes the order in the genus expansion accordingly, and what remains is the result for the appropriate number of punctures. As an example take~(\ref{eq:W1n-special}), which for $\Gamma{=}0$ gives the usual (genus~1) $V_{1,n}$ for purely Neveu-Schwarz case. The~$\Gamma^2$ term instead gives that with two Ramond punctures, there are non-zero {\it genus zero} volumes with $n$ additional NS boundaries that are: \begin{equation}
    V^{(2)}_{0,n}=(-1)^n\frac{(n-1)!}{2}\ ,
\end{equation} precisely the result in Lemma 4.9 of ref.~\cite{norbury2024superweilpeterssonmeasuresmoduli}!  (Again, we are in conventions where there is an overall $(-1)^n$ in front of all of ref.~\cite{norbury2024superweilpeterssonmeasuresmoduli}'s supervolumes.)  Starting instead with formula~(\ref{eq:W2n-special}) the  same kinds of deductions can be made about various other  genera involving Ramond insertions. For example, genus zero and  four Ramond punctures  is read off from the $\Gamma^4$ term, and it gives: 
\begin{equation}
\label{eq:N=1V0n-R4}
    V^{(4)}_{0,n}=
\frac{(-1)^n }{4\cdot 4!}\,
\left[
(2\pi)^2 (n+2)!
+ (n+1)!\sum_{i=1}^{n} b_i^2
\right]\ ,
\end{equation}
  precisely the result  in  ref.~\cite{norbury2024superweilpeterssonmeasuresmoduli}'s Lemma 4.10. The same formula predicts a genus one result for two Ramond punctures with a similar form, except with an additional factor of $\frac52$, which should be interesting to confirm by other methods. 
  
It seems clear from this procedure that the $\Gamma$-dependent formulae derivable using this paper's techniques will yield more closed-form results for volumes with   Ramond insertions at various genus. This is a precise and simple realization of the observations in refs.~\cite{norbury2024superweilpeterssonmeasuresmoduli,Alexandrov:2024kuj} that the Ramond volumes are in a sense deformations of the case of purely Neveu-Schwarz volumes. It's all naturally contained in the string equation~(\ref{eq:big-string-equation}) by just turning on $\Gamma$.\footnote{Note for example that  refs.~\cite{Norbury:2020vyi,norbury2024superweilpeterssonmeasuresmoduli})  derive  modified Virasoro constraints in order to express the deformation picture in terms of recursion. Such constraints  were established using the string equation approach long ago in refs.~\cite{Dalley:1992br,Johnson:1994vk}.}

One could yield these results in a different way, by turning this all into a procedure where  $\hbar\Gamma$ is held fixed in the string equation (yielding a classical leading term in $u_0(x)$  along the same lines as was done for BPS states in the ${\cal N}{=}2$ case in  ref.~\cite{Johnson:2023ofr}) and then a new family of $t_k$, with $\Gamma$ dependence, can be found that yields the volumes we've been reading off here from the general $\Gamma$ results. There are certain advantages to pursuing this methodology, but this and further aspects of this simple approach to volumes with Ramond punctures will all be explained in a subsequent publication~\cite{Johnson:2026jls} since it takes us too far off track in this paper.

So, mechanically we can see why the framework presented in this paper works so nicely, but, {\it really} why is it all working so nicely? Some wisdom from the theory of integrable systems is useful here. At the heart of the matter is the fact that $u(x)$ satisfies the (integrable) KdV flows~(\ref{eq:kdv-flows}), that it satisfies the string equation, and  that it is a potential of a Hamiltonian~(\ref{eq:aux-hamiltonian}).

In writing $u(x) {=} u_0(x){+} O(\hbar^2)$, we’re asking that the $R_k[u]$ have a classical piece, $u_0^k$, and an $O(\hbar^2)$ piece which, in the sense of the  flows being wave equations, corresponds to the dispersive part of the flow, made of multiple derivatives, that normally works in concert with the non-linear classical part to give soliton solutions. The flow looks like:
\begin{equation}
    \frac{\partial u}{\partial t_k}=(k+1)u_0^\prime u_0^{k}+O(\hbar^2)\ ,
\end{equation}
and 
writing things in terms of $u_0$ is equivalent to restricting to the non-dispersive part of the flow. The string equation’s job, in this language, is to organize the dispersive parts of $u(x)$, the $u_{2g}$, at successively higher orders in~$\hbar^2$, in terms of $u_0$. The fact that they are double derivatives of ratios of combinations of $u_0^{\prime}$ and higher derivatives means that $u(x)$ has been organized into a form resembling a ``quasi-Miura'' transformation, part of a framework that has been extensively studied in {\it  e.g.} refs.~\cite{Dubrovin:2001gee,Dubrovin:2019yrl}, where it has been shown that this form is naturally imposed when $F$ is a $\tau$-function satisfying  the Virasoro constraints of ref.~\cite{Dijkgraaf:1991rs,Fukuma:1991jw,David:1990ge} (which are in turn equivalent to the string equation plus the KdV flows).

In that context, what  this paper makes explicit is that the loop operator itself becomes a very simple object, $\delta_{E_i}^{(u_0)}$ in equation~(\ref{eq:delta-u0}), when restricted to just the non-dispersive sector.
Ultimately, the $W_{g,n}$ themselves end up being built out of ratios of derivatives of $u_0$, in particular combinations that are picked out through the action of $\delta_{E_i}^{(u_0)}$.  The particular combinations presumably  have some natural geometrical meaning in the language of ``Frobenius manifolds'' explored in refs.~\cite{Dubrovin:2001gee}, likely connected to the fact that the $W_{g,n}$ must transform in certain ways as the function $u_0(x)$ changes. This seems natural here once it is recalled that $u_0(x)$ is simply a way of parameterizing the physical object itself, the density (see equation~(\ref{eq:integral-representation})), and by extension all the correlators.

What of the role of ODEs?   In the recent work that led to this paper, the centrality of the Gel'fand-Dikki equation~(\ref{eq:gelfand-dikii}) in yielding information about the ${ W}_{g,1}(z)$ was emphasized. An interesting aspect of this is  that the ODE  naturally supplies {\it non-perturbative} data about ${ W}_{g,1}(z)$ that goes beyond the genus expansion. This was explored recently in ref.~\cite{Johnson:2026jbq} by developing transseries ansatz of the equation which allowed the determination of ZZ, FZZT and mixed FZZT-ZZ instanton contributions to the correlators, as well as computation of formulae for the large~$g$ asymptotics of various Weil-Petersson volumes and generalizations thereof.

A pattern suggests itself here. The string equation itself is an ODE supplying non-perturbative data,  pertaining to the free energy (no boundaries). Gel'fand-Dikii pertains to the one boundary case. It seems natural to wonder whether there is an ODE for the two-boundary case, another for three boundaries, and so on.
Such ODEs would seem to hold the key to neatly excavating non-perturbative physics, but it  is unclear at this time what structures to seek. Since the key object in the the one-boundary case, ${\widehat R}(x,E)$, is the diagonal of the integrand of the Darboux-Christoffel kernel, one possibility is that the ODE to seek is one that governs the full kernel (not just the diagonal). Hence, ref.~\cite{Johnson:2025dyb} proposed an object ${\widehat S}(E,E^\prime,x)$, which reduces to ${\widehat R}(E,x)$ when $E^\prime=E$, and a differential equation for it. The idea (if correct) would be that the higher point correlators come from constructing them as determinants based on $\int dx{\widehat S}(E,E^\prime,x)$. How that is to work remains to be seen.

Finally, another  avenue of  exploration is the following: Now that we can construct  explicit formulae for the  $W_{g,n}(u_0,u_0',u_0^{\prime\prime},\cdots)$, it is worth considering the following question.
Just what are the constraints on the functions $u_0(x)$
that can be inserted into them to define  consistent data for a suitable theory? It is natural to wonder if (or perhaps hope that) the allowed class of $u_0(x)$ functions   could be much larger than that which  naturally arises in  matrix models. If so, these formulae could be  useful  beyond the context in which they were first discovered. The links between the integrable systems and intersection theory frameworks (explored in, {\it e.g.,} the aforementioned refs.~\cite{Dubrovin:2001gee,Dubrovin:2019yrl})  are probably a useful starting point to  further explore this interesting  avenue.

\bigskip
\begin{acknowledgments}
CVJ   thanks  the  US Department of Energy (\protect{DE-SC} 0011687) for  support,  and  Amelia for her support and patience.    
\end{acknowledgments}

\bigskip
\appendix

\begin{widetext}
\section{The Free Energy up to genus four}
\label{app:free-energy-corrections}
By recursively solving the string equation, as described in Section~\ref{sec:toolbox},  $u_{2g}(x)$ always comes out as a second derivative of some combination of $u_0(x)$ and its derivatives. This is to be identified as $2F_g$, the genus $g$ contribution to the free energy. (As mentioned in Section~\ref{sec:discussion} (the discussion section) this can be proven in the integrable systems framework by regarding $u_0$ as the dispersionless part of $u(x)$, and the $u_{2g}$ as a dispersive part that amounts to a ``quasi-Miura'' transformation on $u(x)$. See refs.~\cite{Dubrovin:2001gee,Dubrovin:2019yrl}.) The first few  are collected here:
\begin{eqnarray}
    F_1&=&-\frac{1}{24}\ln u_0^\prime\ ,\\ 
   F_2
&=&
\frac{(u_0'')^3}{180\,u_0'^4}
-\frac{7\,u_0^{(3)}u_0''}{960\,u_0'^3}
+\frac{u_0^{(4)}}{576\,u_0'^2}\ ,\\
    F_3&=&
-\frac{u_0^{(7)}}{10368\,u_0'^3}
+\frac{7\,u_0^{(6)}u_0''}{5760\,u_0'^4}
+\frac{53\,u_0^{(5)}u_0^{(3)}}{20160\,u_0'^4}
-\frac{353\,u_0^{(5)}u_0''^2}{40320\,u_0'^5}
+\frac{103\,(u_0^{(4)})^2}{60480\,u_0'^4}
-\frac{1273\,u_0^{(4)}u_0^{(3)}u_0''}{40320\,u_0'^5}
+\frac{83\,u_0^{(4)}u_0''^3}{1890\,u_0'^6}
\nonumber\\
&&\hskip0.5cm\quad
-\frac{59\,(u_0^{(3)})^3}{8064\,u_0'^5}
+\frac{83\,(u_0^{(3)})^2u_0''^2}{896\,u_0'^6}
-\frac{59\,u_0^{(3)}u_0''^4}{378\,u_0'^7}
+\frac{5\,u_0''^6}{81\,u_0'^8}\ .
\end{eqnarray} The algebra becomes increasingly involved at higher order, but can be nicely organized in different ways. For example, a recursion relation for $F_g$ has been known for some time (see {\it e.g.} ref.\cite{Dubrovin:2001gee}), and from a helpful form of it given in equation (5.17) of ref.~\cite{Liu:2024wwu}, an expression for $F_4$ can be readily obtained:
\begin{eqnarray}
\begin{aligned}
F_4
=&
-\frac{u_0^{(10)}}{995328\,u_0'^4}
+\frac{1}{u_0'^5}\left[
\frac{2069}{11612160}\,u_0^{(4)}u_0^{(7)}
+\frac{197}{774144}\,u_0^{(5)}u_0^{(6)}
+\frac{7}{276480}\,u_0''u_0^{(9)}
+\frac{163}{1935360}\,u_0^{(3)}u_0^{(8)}
\right]
\\[1ex]
&\quad
-\frac{1}{u_0'^6}\left[
\frac{229}{124416}\,\bigl(u_0^{(4)}\bigr)^3
+\frac{2323}{6451200}\,\bigl(u_0''\bigr)^2u_0^{(8)}
+\frac{9221}{3870720}\,u_0''\bigl(u_0^{(5)}\bigr)^2
+\frac{59}{21504}\,\bigl(u_0^{(3)}\bigr)^2u_0^{(6)}\right.
\\
&\qquad\qquad\left.
+\frac{20639}{9676800}\,u_0''u_0^{(3)}u_0^{(7)}
+\frac{15179}{3870720}\,u_0''u_0^{(4)}u_0^{(6)}
+\frac{949}{110592}\,u_0^{(3)}u_0^{(4)}u_0^{(5)}
\right]
\\[1ex]
&\quad
+\frac{1}{u_0'^7}\left[
\frac{12035}{193536}\,u_0''\bigl(u_0^{(3)}\bigr)^2u_0^{(5)}
+\frac{44201}{552960}\,u_0''u_0^{(3)}\bigl(u_0^{(4)}\bigr)^2
+\frac{212267}{58060800}\,\bigl(u_0''\bigr)^3u_0^{(7)}\right.
\\
&\qquad\qquad\left.
+\frac{2153}{57600}\,\bigl(u_0^{(3)}\bigr)^3u_0^{(4)}
+\frac{60941}{2150400}\,\bigl(u_0''\bigr)^2u_0^{(3)}u_0^{(6)}
+\frac{171343}{3870720}\,\bigl(u_0''\bigr)^2u_0^{(4)}u_0^{(5)}
\right]
\\[1ex]
&\quad
-\frac{1}{u_0'^8}\left[
\frac{415273}{1658880}\,\bigl(u_0''\bigr)^3u_0^{(3)}u_0^{(5)}
+\frac{13138507}{19353600}\,\bigl(u_0''\bigr)^2\bigl(u_0^{(3)}\bigr)^2u_0^{(4)}\right.
\\
&\qquad\qquad\left.
+\frac{1823}{11340}\,\bigl(u_0''\bigr)^3\bigl(u_0^{(4)}\bigr)^2
+\frac{12907}{453600}\,\bigl(u_0''\bigr)^4u_0^{(6)}
+\frac{22809}{143360}\,u_0''\bigl(u_0^{(3)}\bigr)^4
\right]
\\[1ex]
&\quad
+\frac{1}{u_0'^9}\left[
\frac{101503}{64800}\,\bigl(u_0''\bigr)^4u_0^{(3)}u_0^{(4)}
+\frac{2243}{12960}\,\bigl(u_0''\bigr)^5u_0^{(5)}
+\frac{305129}{207360}\,\bigl(u_0''\bigr)^3\bigl(u_0^{(3)}\bigr)^3
\right]
\\[1ex]
&\quad
-\frac{1}{u_0'^{10}}\left[
\frac{4952}{6075}\,\bigl(u_0''\bigr)^6u_0^{(4)}
+\frac{14903}{4320}\,\bigl(u_0''\bigr)^5\bigl(u_0^{(3)}\bigr)^2
\right]
+\frac{386}{135}\,\frac{\bigl(u_0''\bigr)^7u_0^{(3)}}{u_0'^{11}}
-\frac{926}{1215}\,\frac{\bigl(u_0''\bigr)^9}{u_0'^{12}}\ . \label{eq:F4}
\end{aligned}
\end{eqnarray}
From these all  the basic formulae for ${\widehat W}_{g,n}(E_1,E_2,\cdots,E_n)$ up to $g=4$ can be easily constructed by just acting repeatedly, according to~(\ref{eq:Wgn-E}), with $\delta^{(u_0)}_{E_i}$ given in equation~(\ref{eq:delta-u0}).

\section{Additional examples of the general formulae}
\label{app:more-formulae}

To start with listing some one-boundary cases all together, the first three ${\widehat Q}_g(x,E_1)$ are given below, along with a reminder that they are the ${\widetilde W}_{g,1}(x,E_1)$:
\begin{equation}
    {\widetilde W}_{g,1}(x,E_1)=\delta_{E_1}^{(u_0)}\cdot F_g\equiv \widehat{Q}_g(x,E_1)\ ,
\end{equation}
where:
\begin{eqnarray}
\widehat{Q}_1(x,E_1)=-\frac{u_0^{\prime\prime}(x)}{48 u_0^\prime(x)[u_0(x)-E_1]^{3/2}}+\frac{u_0^\prime(x)}{32[u_0(x)-E_1]^{5/2}}\ ,
\label{eq:queue-one}
\end{eqnarray}
%
%
\begin{eqnarray}
   &&\hskip-1.1cm \widehat{Q}_2(x,E_1)=\frac{105u_0'^3}{2048[u_0-E_1]^{11/2}}-\frac{203u_0'u_0''}{3072[u_0-E_1]^{9/2}}+\frac{29u_0'u_0'''-3u_0''^2}{1536u_0'[u_0-E_1]^{7/2}} \\
        &&\hskip-1.0cm-\frac{17u_0''^3-34u_0'u_0''u_0'''+15u_0'^2u_0^{(4)}}{3840u_0'^3[u_0-E_1]^{5/2}}-\frac{64u_0''^4-111u_0'u_0''^2u_0'''+21u_0'^2u_0'''^2+31u_0'^2u_0''u_0^{(4)}-5u_0'^3u_0^{(5)}}{5760u_0'^5[u_0-E_1]^{3/2}}\ .
        \nonumber
\label{eq:queue-two}
\end{eqnarray}
and:
\begin{equation}
\label{eq:Q3}
    \begin{split}
    &\widehat{Q}_3(x,E_1)=\frac{25025u_0'^5}{65536[u_0-E_1]^{17/2}}-\frac{77077u_0'^3u_0''}{98304[u_0-E_1]^{15/2}}+\frac{11u_0'(633u_0''^2+503u_0'u_0''')}{24576[u_0-E_1]^{13/2}}\\
    & \hskip1.5cm-\frac{57u_0''^3+656u_0'u_0''u_0'''+607u_0'^2u_0^{(4)}}{12288u_0'[u_0-E_1]^{11/2}}\\
    & \hskip1.5cm+\frac{-2385u_0''^4+4740u_0'u_0''^2u_0'''-1315u_0'^2u_0'''^2-1638u_0'^2u_0''u_0^{(4)}+1006u_0'^3u_0^{(5)}}{110592u_0'^3[u_0-E_1]^{9/2}}\\
    &\hskip1.5cm +\frac{\widehat{Q}_{3,3}}{[u_0-E_1]^{7/2}}+\frac{\widehat{Q}_{3,2}}{[u_0-E_1]^{5/2}}+\frac{\widehat{Q}_{3,1}}{[u_0-E_1]^{3/2}}\ , 
    \end{split}
\end{equation}
where:
\begin{equation}
\begin{split}
    &\widehat{Q}_{3,3}=-\frac{14055u_0''^5-34320u_0'u_0''^3u_0'''+11094u_0'^2u_0''^2u_0^{(4)}}{387072u_0'^5}\\
    &\hskip1.5cm-\frac{u_0'^2u_0''\left(16820u_0'''^2-2751u_0'u_0^{(5)}\right)+u_0'^3\left(-5413u_0'''u_0^{(4)}+539u_0'u_0^{(6)}\right)}{387072u_0'^5}\\
\end{split}
\end{equation}
\begin{equation}
\begin{split}
&\widehat{Q}_{3,2}=\frac{-54080u_0''^6+148710u_0'u_0''^4u_0'''-45462u_0'^2u_0''^3u_0^{(4)}}{967680u_0'^7}\\
    &\hskip1.5cm+\frac{9u_0'^2u_0''^2\left(-10965u_0'''^2+1124u_0'u_0^{(5)}\right)+14u_0'^3u_0''\left(2721u_0'''u_0^{(4)}-119u_0'u_0^{(6)}\right)}{967680u_0'^7}\\
    &\hskip1.5cm +\frac{9045u_0'''^3-2483u_0'(u_0^{(4)})^2-3764u_0'u_0'''u_0^{(5)}+175u_0'^2u_0^{(7)}}{967680u_0'^4}\\
\end{split}
\end{equation}
\begin{equation}
\begin{split}
&\widehat{Q}_{3,1}=-\frac{179200u_0''^7-530880u_0'u_0''^5u_0'''+152256u_0'^2u_0''^4u_0^{(4)}}{1451520u_0'^9}\\
    &\hskip1.5cm -\frac{3u_0'^2u_0''^3\left(142750u_0'''^2-10607u_0'u_0^{(5)}\right)+9u_0'^3u_0''^2\left(-19147u_0'''u_0^{(4)}+549u_0'u_0^{(6)}\right)}{1451520u_0'^9}\\
    &\hskip1.5cm -\frac{3u_0''\left[-26835u_0'''^3+7209u_0'u_0'''u_0^{(5)}+u_0'\left(4643(u_0^{(4)})^2-182u_0'u_0^{(7)}\right)\right]}{1451520u_0'^6}\\
    &\hskip1.5cm -\frac{\left[19422u_0'''^2u_0^{(4)}+35u_0'^2u_0^{(8)}-15u_0'\left(146u_0^{(4)}u_0^{(5)}+93u_0^{(3)}u_0^{(6)}\right)\right]}{1451520u_0'^5}\ .
\end{split}
\end{equation}

In the main text, $W_{1,1}(z_1)$ and $W_{1,2}(z_1,z_2)$ and and their tilded counterparts with $E_i$ dependence were discussed extensively. Recall that the latter can be obtained from the former by restoring the Jacobian factors $\prod_i^n(-2z_i)$, and writing $z_i=(u_0(\mu)-E_i)^\frac12$. Here is $W_{1,3}(z_1,z_2,z_3)$ along with the $\ell_i$ dependent counterpart to ${\widetilde W}_{1,3}(E_1,E_2,E_3)$:
\begin{equation}
\label{eq:W13}
\begin{aligned}
W_{1,3}(z_1,z_2,z_3)
=
\frac{1}{z_1^2 z_2^2 z_3^2}\Bigg[
& A_{13}(\mu)
+ B_{13}(\mu)\sum_{i=1}^3 \frac{1}{z_i^2}
+ M_{13}(\mu)\!\left(
10\sum_{i=1}^3 \frac{1}{z_i^4}
+9\sum_{1\le i<j\le 3}\frac{1}{z_i^2 z_j^2}
\right)
\\[4pt]
&+ N_{13}(\mu)\!\left(
35\sum_{i=1}^3 \frac{1}{z_i^6}
+30\sum_{i\ne j}\frac{1}{z_i^4 z_j^2}
+18\,\frac{1}{z_1^2 z_2^2 z_3^2}
\right)
\Bigg],
\end{aligned}
\end{equation}
with
\begin{equation}
A_{13}(\mu)
=
\frac{u_0^{(4)}(\mu)}{24\,u_0'(\mu)}
-\frac{u_0''(\mu)u_0^{(3)}(\mu)}{8\,u_0'(\mu)^2}
+\frac{u_0''(\mu)^3}{12\,u_0'(\mu)^3}\ , \qquad
B_{13}(\mu)
=
-\frac{3}{16}\,u_0^{(3)}(\mu)
+\frac{u_0''(\mu)^2}{16\,u_0'(\mu)},
\end{equation}
\begin{equation}
M_{13}(\mu)=\frac{1}{16}\,u_0'(\mu)u_0''(\mu)\ ,
\qquad
N_{13}(\mu)=-\frac{1}{64}\,u_0'(\mu)^3.
\end{equation}
and:
    \begin{eqnarray}
        &&{\widetilde W}_{1,3}(\ell_1,\ell_2,\ell_3)=-\frac{\sqrt{\ell_1\ell_2\ell_3}}{12\pi^{3/2}} 
e^{-u_0(x)(\ell_1+\ell_2+\ell_3)}
\Bigg[
\frac{u_0^{(4)}(x)}{2 u_0'(x)}
+\left(\frac{u_0''(x)}{u_0'(x)}\right)^3
-\frac{3}{2}\frac{u_0'''(x)}{u_0'(x)}
\left(
\frac{u_0''(x)}{u_0'(x)}+u_0'(x)(\ell_1+\ell_2+\ell_3)
\right) \nonumber
\\
&&\quad
+\frac{1}{2}u_0''(x)^2(\ell_1+\ell_2+\ell_3)
+2u_0'(x)^2u_0''(x)
\Big((\ell_1+\ell_2+\ell_3)^2-(\ell_1\ell_2+\ell_1\ell_3+\ell_2\ell_3)\Big) \nonumber
\\
&&\quad
-\frac{1}{2}u_0'(x)^4
\Big((\ell_1+\ell_2+\ell_3)^3
-2(\ell_1+\ell_2+\ell_3)(\ell_1\ell_2+\ell_1\ell_3+\ell_2\ell_3)\Big)
\Bigg].
    \end{eqnarray}

    In the main text, $W_{2,2}(z_1,z_2)$ was listed and discussed (see equations~(\ref{eq:W22-z-A},\ref{eq:W22-z-B})). 
    Many more examples were generated in the course of this project, but it seems pointless to list the resulting vast formulae here when they can be more easily (and usefully) generated by implementing the simple operator~(\ref{eq:delta-u0}) in a symbolic computation program.

\end{widetext}

\bibliographystyle{apsrev4-1}
\bibliography{references}


\end{document}